\pdfoutput=1
\documentclass{article}

\usepackage{microtype}
\usepackage{graphicx}
\usepackage{booktabs}
\usepackage{amsmath,amssymb,amsfonts,amsthm,mathtools}
\usepackage{pifont}
\newcommand{\cmark}{\ding{51}}
\newcommand{\xmark}{\ding{55}}
\usepackage{xcolor}
\usepackage{url}
\usepackage{xurl}
\usepackage{tabularx}
\usepackage{hyperref}
\newcommand{\supp}[1]{Appendix~\ref{#1}}

\usepackage[accepted]{mlsys2025}
\makeatletter\renewcommand{\Notice@String}{Preprint. DOI: \href{https://doi.org/10.5281/zenodo.23674584}{10.5281/zenodo.23674584}.\\ Code: \url{https://github.com/hpdic/libcertex}.\\ Copyright \copyright{} 2026 Dongfang Zhao.}\renewcommand{\mlsysauthor}[2]{\mbox{\bf #1} ({\normalfont #2})\addtofullauthorlist{#1}}\renewcommand{\printAffiliationsAndNotice}[1]{{\let\thefootnote\relax\long\def\@makefntext##1{\noindent##1}\footnotetext{\Notice@String}}}\makeatother

\makeatletter
\long\def\@makecaption#1#2{
 \def\@tempa{table}\ifx\@captype\@tempa\vskip 4pt\else\vskip -8pt\fi
        \baselineskip 11pt
        \setbox\@tempboxa\hbox{#1. #2}
        \ifdim \wd\@tempboxa >\hsize
        \sbox{\newcaptionbox}{\small\sl #1.~}
        \newcaptionboxwid=\wd\newcaptionbox
        \usebox\newcaptionbox {\footnotesize #2}
        \else
          \centerline{{\small\sl #1.} {\small #2}}
        \fi}
\makeatother

\hypersetup{hypertexnames=false}

\newcommand{\sysname}{\textsc{Zarc}}

\newcommand{\lift}{\ell}
\newtheorem{proposition}{Proposition}

\graphicspath{{figures/}}
\mlsystitlerunning{Zarc: Exact Nearest Neighbors over a Billion Vectors in Seconds}

\begin{document}

\twocolumn[
\mlsystitle{\sysname: Exact Nearest Neighbors over a Billion Vectors in Seconds}

\begin{mlsysauthorlist}
\mlsysauthor{Dongfang Zhao}{dzhao@uw.edu}
\end{mlsysauthorlist}

\mlsyskeywords{exact nearest-neighbor search, vector search, GPU, tensor cores, LLM attention}

\vskip 0.3in
]

\printAffiliationsAndNotice{}

\begin{abstract}
Exact $k$-nearest-neighbor vector search lies on the critical path of machine learning systems, from sparse attention in the latest large language models (LLMs) to long-context memories and training-data deduplication. Exhaustive GPU search scans the whole dataset for every query, while approximate indexes return inexact neighbors. Recent methods skip almost every vector with a cheap half-precision proof that it lies too far away, but only on one GPU that holds the whole dataset. Our insight is to keep only the short rows of these proofs in GPU memory and to fetch the few remaining full vectors from host memory. Our system, \sysname, answers Euclidean, cosine, inner-product, and filtered queries exactly under updates. On four H100 GPUs, it answers 10,000 exact queries over one billion vectors in 1.1~seconds instead of five minutes. On eleven open LLMs, it retrieves the exact top-10 attention keys 2.2$\times$ to 13.1$\times$ faster than exhaustive search. In sparse attention over one million tokens, it runs 3.1$\times$ faster than bf16 selection with FlashInfer, which misses the exact top-64 for most queries.
\end{abstract}
\section{Introduction}
\label{sec:intro}

Exact $k$-nearest-neighbor vector search lies on the critical path of machine learning systems, in which every missed neighbor becomes an error of the model or of its data. In large language models (LLMs), sparse attention such as that of DeepSeek-V4~\cite{deepseek2026v4} attends only to the top-$k$ entries of the preceding context for each query token. Long-context memories such as kNN-LM~\cite{khandelwal2020knnlm} and Memorizing Transformers~\cite{wu2022memorizing} likewise attend to the nearest keys of a memory that grows with every generated token. Beyond attention, semantic deduplication such as SemDeDup~\cite{abbas2023semdedup} relies on nearest neighbors, as does the evaluation of representations such as CLIP~\cite{radford2021clip} and DINOv2~\cite{oquab2024dinov2} and of approximate search itself~\cite{aumuller2020annbench,simhadri2022bigann}. These workloads run in batches over hundreds of millions to billions of vectors, which change over time and carry the metadata of filters.

\begin{figure*}[t]
\centering
\includegraphics[width=\textwidth]{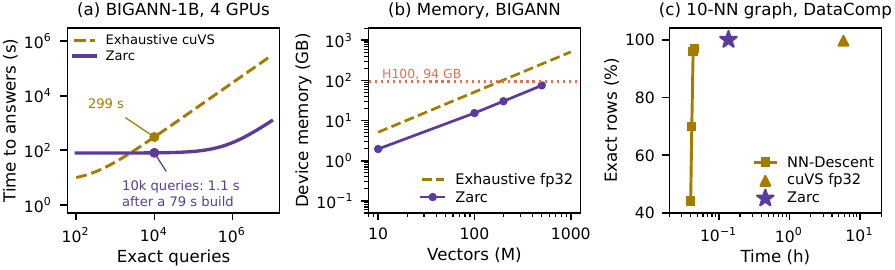}
\caption{Time, memory, and exactness costs of exact search, and the effect of \sysname.}
\label{fig:opener}
\end{figure*}

GPU vector search systems either evaluate every distance or search an approximate index. Exhaustive search in Faiss~\cite{johnson2021faiss} and cuVS~\cite{nvidia2025cuvs} scans every vector for every query. As a result, 10,000 queries over BIGANN-1B~\cite{jegou2011bigann} take about five minutes on four H100 GPUs (Figure~\ref{fig:opener}(a)). Exhaustive search also keeps the raw vectors in device memory, which model weights and KV caches claim as well. One billion 128-dimensional vectors alone occupy 512~GB in single precision (Figure~\ref{fig:opener}(b)). Approximate indexes such as HNSW~\cite{malkov2020hnsw}, DiskANN~\cite{subramanya2019diskann}, CAGRA~\cite{ootomo2024cagra}, and product-quantized inverted files~\cite{jegou2011pq,babenko2012imi} return inexact neighbors without a per-query guarantee, whose errors reach the model and its data. For instance, CAGRA at its most accurate setting returns the exact top-10 attention keys of Qwen2.5-7B~\cite{yang2024qwen25} for at most 45\% of the queries. Similarly, NN-Descent~\cite{dong2011nndescent} leaves 3\% to 56\% of the rows of the DataComp~\cite{gadre2023datacomp} deduplication graph inexact (Figure~\ref{fig:opener}(c)). On BIGANN-10M at 0.1\% selectivity, the filtered graph index ACORN~\cite{patel2024acorn} returns the exact neighbors of only 3.4\% of the queries.

The two designs differ in the evidence on which they skip a vector. Exhaustive search skips no vector and therefore pays for every distance, whereas approximate search skips vectors without proof that they lie beyond the $k$-th distance and therefore loses neighbors. A vector can be skipped without any loss of exactness only with such a proof, namely a certified lower bound on its distance above the $k$-th distance. Certified exclusion~\cite{zhao2026certex} obtains this proof at the speed of the tensor cores by computing the lower bounds in half precision and including the rounding error of the product in every bound. As a result, only the few vectors that are not skipped need exact distances. Yet the CertEx engine that introduced certified exclusion holds the whole dataset in the memory of one GPU and supports no deletions, concurrent updates, filters, or inner products.

These proofs read only a short half-precision row of every vector. Our insight is to keep only these rows in GPU memory and the full-precision vectors in host memory, because exact verification reads the full vectors of only the few vectors that remain. This paper turns the above insight into \sysname, an exact vector search system with the interface of cuVS~\cite{nvidia2025cuvs} that installs as a Python package. \sysname\ derives tight upper bounds on the $k$-th distance from partitions of the dataset and verifies the survivors from host memory in the order of their certified lower bounds, which keeps every answer exact under insertions, deletions, concurrent queries, filters, and inner-product queries.

We evaluate \sysname\ on datasets of up to one billion vectors and on the attention keys of eleven open LLMs, against cuVS~\cite{nvidia2025cuvs} brute force, CAGRA~\cite{ootomo2024cagra}, IVF-PQ~\cite{jegou2011pq}, DiskANN~\cite{subramanya2019diskann}, and filtered graph indexes. Figure~\ref{fig:opener} shows the effect of \sysname\ on the time, the memory, and the exactness of exact search. On four H100 GPUs, \sysname\ indexes BIGANN-1B~\cite{jegou2011bigann} in 79~s and answers the same 10,000 queries exactly in 1.1~s (Figure~\ref{fig:opener}(a)). On the eleven LLMs, from 7B to 1.6T parameters, \sysname\ returns the exact top-10 attention keys at 2.2$\times$ to 13.1$\times$ the throughput of exhaustive cuVS search, while the bf16 matrix products of LLM code miss the exact top-10 for 14\% to 86\% of the queries (Figure~\ref{fig:eval_llm}). In top-$k$ sparse attention over a one-million-token context, exact selection by \sysname\ shortens the forward pass of Qwen2.5-7B-Instruct-1M~\cite{yang2025qwen251m} by 3.1$\times$ against bf16 selection with the top-$k$ kernel of FlashInfer~\cite{ye2025flashinfer} and below the time of dense attention (Figure~\ref{fig:eval_llm_e2e}). On one H100, \sysname\ searches BIGANN-500M exactly within 71~GB of device memory, against 256~GB for exhaustive search (Figure~\ref{fig:opener}(b)). \sysname\ also runs 692$\times$ faster than filtered CAGRA at 0.1\% selectivity and builds the exact 10-nearest-neighbor graph of DataComp-12.8M~\cite{gadre2023datacomp} 42$\times$ faster than exhaustive search (Figure~\ref{fig:opener}(c)).

This paper makes the following contributions:
\begin{itemize}
  \item We design certified screening across the memory hierarchy, with half-precision screening rows in GPU memory, full-precision vectors in host memory, and verification of the survivors in the order of their certified lower bounds. Thresholds derived from data partitions and projections fitted to each partition cut the raw vectors that a query reads over PCIe to 699 on BIGANN-500M~\cite{jegou2011bigann}. Updates, filters, and inner products reuse the same certified bounds.
  \item We implement \sysname\ as a library with the interface of cuVS~\cite{nvidia2025cuvs}, ahead-of-time compiled tensor-core kernels for Hopper and Ampere GPUs, concurrent updates, a calibrated cost model, and sharded search over the GPUs of one server.
  \item We evaluate \sysname\ on eleven open LLMs from 7B to 1.6T parameters, in sparse attention over a one-million-token context, and on billion-scale benchmarks, filtered search, streaming updates, semantic deduplication, and passage retrieval, where it is exact for every query, up to 270$\times$ faster than exhaustive GPU search, and 2.2$\times$ to 13.1$\times$ faster on LLM attention keys.
\end{itemize}

\section{Motivation and Challenges}
\label{sec:background}

\subsection{Exact Search Workloads}
\label{sec:bg:exact_ann}
Various critical components of modern AI systems rely on efficient and accurate vector search capabilities,
such as sparse attention, long-context memories, and deduplication. These applications need exact answers over large datasets that change over time. Exactness requires more than single-precision distances whenever the largest scores of a query lie in near ties, which attention scores often do (Section~\ref{sec:eval:llm}). For example, the ten largest attention scores of a GLM-5.3~\cite{glm2026glm5} query lie within 0.03\% of each other. The datasets are large enough that their raw vectors compete for device memory with model weights and KV caches. For example, BIGANN-500M~\cite{jegou2011bigann} alone occupies 256~GB in single precision. The datasets also change while queries run, which rules out rebuilding the index for every change.

\begin{table}[t]
\centering
\resizebox{\columnwidth}{!}{
\begin{tabular}{lcccc}
\toprule
System & Exact & Beyond HBM & Updates & Exact filters \\
\midrule
cuVS brute force~\cite{nvidia2025cuvs} & partial & \xmark & \cmark & partial \\
cuVS IVF-PQ~\cite{jegou2011pq} & \xmark & \cmark & partial & \xmark \\
CAGRA~\cite{ootomo2024cagra} & \xmark & \xmark & partial & \xmark \\
DiskANN~\cite{subramanya2019diskann} & \xmark & \cmark & \xmark & \xmark \\
FreshDiskANN~\cite{singh2021freshdiskann} & \xmark & \cmark & \cmark & \xmark \\
ACORN~\cite{patel2024acorn} & \xmark & \xmark & partial & \xmark \\
Filtered-DiskANN~\cite{golub2024filtereddiskann} & \xmark & \cmark & \xmark & \xmark \\
CertEx~\cite{zhao2026certex} & \cmark & \xmark & partial & \xmark \\
\sysname & \cmark & \cmark & \cmark & \cmark \\
\bottomrule
\end{tabular}
}
\caption{Properties of GPU and CPU vector search systems.}
\label{tab:systems}
\end{table}

\begin{figure}[t]
\centering
\includegraphics[width=\columnwidth]{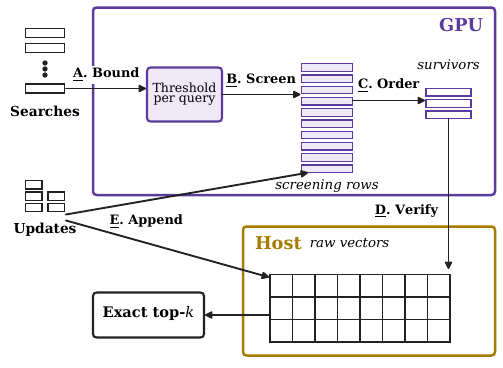}
\caption{\sysname: search path (A to D) and updates (E).}
\label{fig:arch}
\end{figure}

The queries also vary in form and arrive in batches. A query may restrict its neighbors to a metadata label or rank them by inner product, as dense retrievers such as TAS-B~\cite{hofstatter2021tasb} do. Deduplication, benchmark evaluation, and the decoding batches of LLM serving systems such as Orca~\cite{yu2022orca} and vLLM~\cite{kwon2023vllm} submit many queries at once. One pass over the index can thus serve a whole batch.

\subsection{Comparison of Existing Systems}
\label{sec:bg:filtered}

Table~\ref{tab:systems} compares vector search systems on these properties, where a partial mark denotes single-precision exactness or insertions without deletions. Exhaustive search in Faiss~\cite{johnson2021faiss,douze2024faisslib} and cuVS~\cite{nvidia2025cuvs} is exact but keeps the database in device memory and scans it for every query. Inverted files~\cite{babenko2012imi} and proximity graphs such as HNSW~\cite{malkov2020hnsw} avoid the full scan by searching a fraction of the dataset, which makes their recall depend on the probe count or the search width. A graph also absorbs a change of the dataset only by an extension or a rebuild. For instance, a rebuild takes 27.6~s for CAGRA~\cite{ootomo2024cagra} over 8.8M passages on an H100 and 49 minutes for ACORN-$\gamma$~\cite{patel2024acorn} over the 10M vectors of BIGANN-10M~\cite{jegou2011bigann}.

Filters raise the cost of both graph indexes and exhaustive search. At low selectivity, the matching vertices of a graph index often form a disconnected subgraph. Filtered graph indexes such as ACORN~\cite{patel2024acorn} and Filtered-DiskANN~\cite{golub2024filtereddiskann} therefore add predicate-agnostic or label-aware edges and still return approximate answers. Exhaustive cuVS~\cite{nvidia2025cuvs} search stays exact with a bitmap prefilter but still pays for the full scan.

\subsection{Certified Exclusion and Its Limits}
\label{sec:bg:certex}
\label{sec:bg:challenges}

CertEx~\cite{zhao2026certex} skips a vector when a lower bound on its distance to the query exceeds a threshold, which bounds the distance of the $k$-th nearest neighbor from above. The threshold is the $k$-th smallest exact distance from the query to a uniform sample of stored vectors, which CertEx calls anchors. The lower bound needs only a short screening row per vector, which holds in half precision the projection of the vector onto the leading principal components of the dataset. One tensor-core product evaluates the bounds of all screening rows for a query. Because every bound includes the worst-case rounding error of this product, no skipped vector can be a true neighbor. Exact distances in full precision are then computed only for the few vectors that pass this screen, which CertEx calls survivors.

Production workloads expose five limits of CertEx~\cite{zhao2026certex}, which keeps the raw vectors and the screening rows of the whole dataset on one GPU. First, the uniform sample loosens the threshold on clustered and near-duplicate data (Section~\ref{sec:design:ivf}). Second, the raw vectors of a billion-scale dataset do not fit beside its screening rows (Section~\ref{sec:design:tiered_mem}). Third, updates must keep every bound valid without a rebuild (Section~\ref{sec:design:updates}). Fourth, the throughput of the exact configurations varies by up to 5$\times$, which calls for automatic tuning of the number of projected dimensions and of the number of anchors (Section~\ref{sec:design:tuning}). Finally, filters and inner products need certified bounds of their own (Section~\ref{sec:design:filtered}).

\section{Design}
\label{sec:design}

\subsection{System Architecture Overview}
\label{sec:design:overview}

Figure~\ref{fig:arch} shows how \sysname\ executes a batch of queries in four stages on the GPU while the raw vectors stay in host memory. The threshold stage (A) computes for every query a threshold on its $k$-th distance from the exact distances of sampled anchors or of candidates in the nearest partitions of the query. The screening stage (B) then screens the half-precision rows of all vectors on the tensor cores and keeps every row whose certified lower bound does not exceed the threshold of the query (\supp{supp:screen}). Each surviving query-vector pair (C) carries its computed product, from which \sysname\ derives a certified lower bound and orders the survivors of each query. Verification (D) finally reads the raw vectors of the survivors from host memory in this order and computes their exact distances in double precision, until the next lower bound exceeds the $k$-th distance found. Updates (E) append the screening rows and raw vectors of new vectors and mark the rows of deleted vectors, while queries keep running on a consistent view of the index (Section~\ref{sec:design:updates}).

Because every exclusion is certified and every survivor that might be a neighbor is verified, the pipeline returns the exact top-$k$. This guarantee rests on two invariants, namely that every excluded row has a certified lower bound above its threshold and that every threshold bounds the $k$-th distance from above (\supp{supp:correct}). Every mechanism below keeps both invariants, including row placement across memories and GPUs and concurrent updates.

\sysname\ screens on a persistent tensor-core kernel that never writes the product matrix to memory. Each thread block keeps a tile of query rows in shared memory and streams tiles of screening rows through a pipeline of asynchronous loads, whose products accumulate in single precision. Each block reduces its rows in one sequential loop without the split-K schedules of standard matrix-multiply libraries, since the rounding-error bound holds only for this accumulation order. The epilogue then compares every product with the threshold of its query and compacts the survivors without global synchronization. Each thread marks its survivors in a bit mask and reserves output space with one shared-memory atomic. As a result, an empty tile costs one comparison per product and no memory traffic (\supp{sec:impl:kernels}).

\subsection{Tighter Bounds from Partitions}
\label{sec:design:ivf}

\sysname\ tightens the threshold with a $k$-means partition of the dataset, whose partitions it stores as contiguous blocks of rows. For every query, \sysname\ first screens under the anchor threshold of CertEx~\cite{zhao2026certex} only the rows of the few partitions whose centroids are nearest to the query. A routed variant of the screening kernel performs this step in one launch over a list of work items, each of which pairs a tile of queries with a tile of rows from one partition. Among the screened rows, \sysname\ computes the exact distances of a small multiple of $k$ candidates with the smallest certified lower bounds. The $k$-th smallest of these distances gives a second threshold, which we call the partition threshold. Screening of all rows then runs under the smaller threshold.

The partition threshold is safe for any choice of partitions and candidates, because it is an exact distance to $k$ distinct stored vectors and thus bounds the true $k$-th distance from above. The partitions therefore change only the speed of \sysname, whereas the partitions of an inverted file decide which vectors a query can return. Deduplication runs a self-join, in which every stored vector is a query. Such a self-join skips the screening of the nearest partitions. Instead, the queries of a batch in storage order come from one or two partitions, which directly give the threshold of the batch.

\label{sec:design:local_bounds}
On a clustered dataset, the vectors lie far from one global principal subspace but close to the subspace of their own partition. \sysname\ therefore gives every partition its own local projection onto the leading eigenvectors of the covariance of the partition around its centroid, which tightens the bound. Each partition encodes its rows with its own projection, scale, and rounding-error constants, so that the certified bound holds within the partition (\supp{supp:local}). A search projects every query once per partition with one batched matrix product, after which one routed launch screens every partition with its query rows. The anchors and later insertions keep the global projection and a plain launch. Every vector thus meets one certified test, never two.

\subsection{Tiered Placement Across GPUs}
\label{sec:design:tiered_mem}

\sysname\ keeps in GPU memory only the half-precision screening rows and the anchor sample, which every query reads, and keeps the full-precision vectors in pinned host memory, which the GPU reads by zero-copy access. With rows several times shorter than full-precision vectors, the same device memory holds several times as many vectors, and optional one-byte integer rows halve the rows again (\supp{supp:int8}). Host memory, however, sits behind PCIe, whose bandwidth lies more than an order of magnitude below that of device memory. With the survivors at scattered positions, verification must read as few raw vectors as possible.

\sysname\ orders verification by the certified lower bounds themselves. After screening, a kernel computes the certified lower bound of every survivor from its stored product and packs the query and the bound into one integer key. A radix sort of these keys then groups the survivors by query in increasing order of their lower bounds. Verification runs in rounds over all queries of a batch, in which every unfinished query takes a window of its next survivors whose lower bounds lie below its current $k$-th exact distance. The window doubles from round to round instead of assigning one warp to each query, which would serialize the long survivor lists of the hardest queries over PCIe. One flat launch computes the exact distances of a round in double precision with several row reads of each warp in flight (\supp{sec:impl:memory}), after which a second launch merges them into the top-$k$ of each query. A query finishes once its window reaches a lower bound above its $k$-th distance, which limits the reads from host memory to a fraction of the survivors (\supp{supp:ordered}).

\label{sec:design:sharding}
A dataset whose screening rows exceed the memory of one GPU splits into contiguous shards with one index per GPU, which \sysname\ builds in parallel. Shards that screen against their own $k$-th distances would keep many more survivors. \sysname\ instead searches in two rounds. In the first round, every shard computes a threshold for each query from its own anchors. The smallest of these shard thresholds bounds the global $k$-th distance, as any $k$ distinct vectors of one shard do (\supp{supp:shards}). In the second round, every shard screens and verifies under this shared threshold, after which the host merges the exact shard answers. Only one threshold and $k$ pairs per shard and query cross the interconnect.

\subsection{Updates Without Re-Indexing}
\label{sec:design:updates}

\sysname\ inserts and deletes vectors while queries run, without a rebuild. The host and device buffers of the raw vectors and the screening rows are preallocated to a capacity, into which an insertion appends its rows in place. The insertion projects the new vectors with the existing projection, since the lower bound stays valid for vectors that the projection was not computed from. The rounding-error bound of every query, however, depends on the largest norm and error terms among all stored rows. A new vector with a larger norm would therefore invalidate the bound of every running query, unless the bound covered the new rows first.

\sysname\ enforces this order with a publication protocol (\supp{sec:impl:api}). An update writes its rows beyond the published rows without a lock and then publishes the new row count together with the raised error bound under an exclusive lock. Every search batch holds the same lock in shared mode. Each batch thus reads one consistent view and never takes its threshold from an anchor that it does not screen. A deletion marks the row of the deleted vector so that it never survives screening. It also removes the vector from the anchor sample. As insertions keep the projection fixed, \sysname\ periodically recomputes it and rebuilds the rows.

\subsection{Tuning by a Calibrated Cost Model}
\label{sec:design:tuning}

\sysname\ selects for every partition the number of projected dimensions, the vectors per anchor, the query batch size, and the memory tier of the full-precision vectors with a cost model. It is calibrated once per device by microbenchmarks and predicts the time of a batch from the screened rows, the anchor distances, and the expected survivors per query (\supp{sec:eval:tuning}). As every configuration keeps every bound valid, a wrong prediction costs only speed.

The survivors per query depend on the data in a way that no static rule captures (\supp{supp:tuning}). For example, the best number of dimensions shrinks as the dataset grows and depends on the query distribution. Hence, the tuner runs the search on two small prefixes of the dataset and extrapolates the survivors and verified rows per query to the full size by a power law. The cost model then ranks the configurations by their predicted batch time and keeps the fastest one.

\subsection{Filtered and Inner-Product Queries}
\label{sec:design:filtered}

A filtered query restricts its neighbors to the vectors that carry a given label, such as a tenant or a category. Screening needs no connectivity among the matching vectors, which reduces a filter to the removal of rows from the screen. Only the threshold depends on the label, because a threshold from vectors of other labels can lie below the $k$-th distance among the matching vectors and thus exclude true neighbors.

\sysname\ supports one label per vector and one label per query in two ways. In the first way, the epilogue of the screening kernel never keeps a row of another label. The anchors of a query then come only from its label. This way serves any per-row predicate, such as the bitset filters of cuVS~\cite{nvidia2025cuvs}, which only reject rows and leave every other bound unchanged. In the second way, \sysname\ stores every label as a contiguous block of rows with its own anchors and sorts each batch of queries by label. The routed kernel then pairs every tile of queries of one label only with the rows of that label, so that the work of the routed kernel shrinks in proportion to the selectivity (\supp{supp:labels}).

\label{sec:design:mips}
For inner-product queries, \sysname\ bounds the inner product directly. The standard reduction to Euclidean search appends to every vector a coordinate that raises its norm to the largest norm of the dataset. Rounding this coordinate, however, perturbs the norms by up to one unit in the last place and can reorder near ties. \sysname\ instead builds the screening rows without subtracting the mean of the dataset, which preserves every norm. The same tensor-core product then bounds the inner product of every vector from above (\supp{supp:correct}). Every vector whose bound falls below the $k$-th largest exact inner product of any $k$ distinct vectors therefore lies outside the top-$k$. The threshold comes from the exact inner products of the anchors. Verification then ranks the survivors by exact inner products in double precision. Cosine queries screen normalized copies with a slack for their rounding (\supp{supp:cosine}).

\section{Evaluation}
\label{sec:evaluation}

\label{sec:implementation}
We implement \sysname\ in about 15,000 lines of CUDA C++ and Python with tests and benchmarks (\supp{supp:impl}). Table~\ref{tab:api} lists its interfaces, each of which keeps the calls of an existing library, from the index modules of cuVS~\cite{nvidia2025cuvs} and the flat indexes of Faiss~\cite{johnson2021faiss} to the top-$k$ selection of PyTorch~\cite{paszke2019pytorch} and FlashInfer~\cite{ye2025flashinfer}. Moving code to exact search therefore changes little more than its imports. The Python binding also takes PyTorch, CuPy, and NumPy arrays without copies. To compute the partition threshold, \sysname\ screens the two nearest partitions and verifies four candidates per requested neighbor, which touches about 0.2\% of the rows. Its tensor-core kernels are compiled ahead of time for Hopper and Ampere GPUs, so that no query waits for kernel tuning when the batch shape changes. In contrast, the Triton~\cite{tillet2019triton} engine of CertEx~\cite{zhao2026certex} tunes each kernel at run time while the queries wait.

\begin{figure*}[t]
\centering
\includegraphics[width=\textwidth]{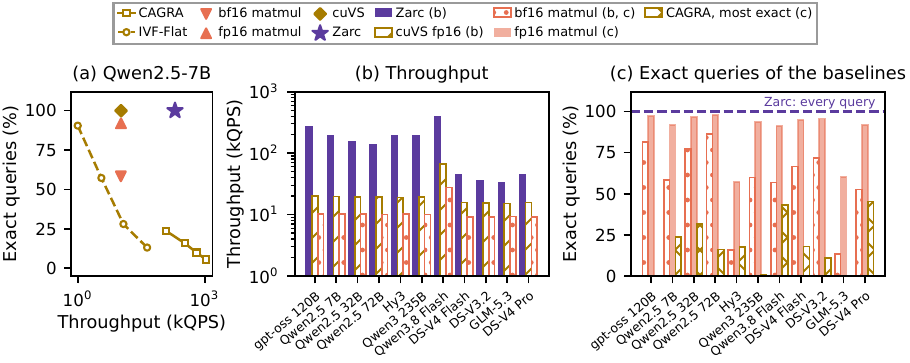}
\caption{Exact top-10 attention keys on eleven open LLMs, with one H100 per KV head.}
\label{fig:eval_llm}
\end{figure*}

\begin{table}[t]
\centering
\resizebox{\columnwidth}{!}{
\begin{tabular}{lll}
\toprule
Library & Drop-in for & Use case \\
\midrule
cuVS, C++ & index modules (\texttt{build}, \texttt{search}) & Vector databases \\
cuVS, C & \texttt{cuvsBruteForce} calls & Vector databases \\
cuVS, Python & \texttt{brute\_force} module & RAG, training-data deduplication \\
Faiss & \texttt{IndexFlatL2}, \texttt{IndexFlatIP} & RAG, benchmark ground truth \\
PyTorch & \texttt{torch.topk} over attention scores & Sparse attention, long-context memory \\
FlashInfer & top-$k$ page-table and ragged transforms & Sparse attention in LLM serving \\
\bottomrule
\end{tabular}
}
\caption{Interfaces of \sysname.}
\label{tab:api}
\end{table}

\begin{figure}[t]
\centering
\includegraphics[width=\columnwidth]{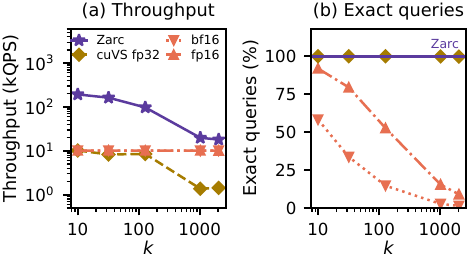}
\caption{Exact top-$k$ on Qwen2.5-7B against $k$.}
\label{fig:eval_llm_k}
\end{figure}

\subsection{Experimental Setup}
\label{sec:eval:setup}

\begin{figure}[t]
\centering
\includegraphics[width=\columnwidth]{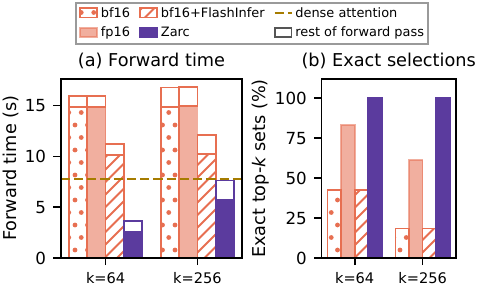}
\caption{Sparse attention end to end over one million tokens.}
\label{fig:eval_llm_e2e}
\end{figure}

\paragraph{Open LLMs.}
The LLM workload covers eleven open LLMs from five developers with 7B to 1.6T parameters in the DeepSeek-V4~\cite{deepseek2026v4,deepseek2025v32}, Qwen~\cite{yang2024qwen25,yang2025qwen3,qwen2026qwen38}, gpt-oss~\cite{openai2025gptoss}, GLM~\cite{glm2026glm5}, and Hy3~\cite{tencent2026hy3} series. Per model, the memory holds the keys of one layer before rotary position embedding (RoPE)~\cite{su2024roformer}, over the train books of PG-19~\cite{rae2020pg19}, with 10M keys for each of the first four KV heads on one H100. Each head receives 10,000 decoding queries from the test books. For the multi-head latent attention~\cite{deepseek2024v2} of DeepSeek and GLM, the memory holds the absorbed latent key that all heads share, whose inner product with the absorbed query of a head equals its attention logit before RoPE. The memory excludes the first four positions of every chunk, the attention sinks~\cite{xiao2024streamingllm} that a deployed model keeps resident. Ground truth comes from fp32 distances to all keys, rescored in fp64 and checked against their rounding error (\supp{supp:llm}).

\paragraph{Platform.}
The main platform is a server with four 94~GB H100 GPUs connected pairwise by NVLink, 192 cores of the Intel Xeon Platinum 8468, and 960~GB of host DRAM, which runs the billion-scale and LLM experiments. A second server with one 94~GB H100 over PCIe~5.0 x16, 48 cores of the same processor, and 240~GB of host DRAM runs the other experiments on one GPU. Both run Ubuntu 24.04, CUDA 12.6 with driver 560.35.05, CUTLASS 3.9.2, PyTorch 2.14.1, and cuVS 26.10. On an A100 of the earlier Ampere generation, every query stays exact and \sysname\ runs 11$\times$ to 26$\times$ faster than exhaustive search (\supp{supp:gpus}).

\paragraph{Datasets.}
Vector workloads use SIFT1M and BIGANN~\cite{jegou2011bigann} of SIFT descriptors~\cite{lowe2004sift}, DINOv2~\cite{oquab2024dinov2} embeddings of ImageNet~\cite{deng2009imagenet}, CLIP~\cite{radford2021clip} embeddings of DataComp~\cite{gadre2023datacomp}, Deep1B~\cite{babenko2016deep}, and two embeddings of MS MARCO~\cite{nguyen2016msmarco} for cosine and inner-product retrieval. BIGANN and Deep1B reach one billion vectors, while DINOv2 and DataComp contain the clustered and near-duplicate data on which uniform thresholds loosen. Every dataset comes with exact ground truth for all queries, computed in double precision (\supp{supp:eval}).

\paragraph{Baselines.}
Exhaustive search runs cuVS~\cite{nvidia2025cuvs} brute force in single precision, with a bitmap prefilter or a search over only the matching vectors for filtered queries. The approximate GPU baselines from cuVS are CAGRA~\cite{ootomo2024cagra} with a bitset filter, IVF-PQ~\cite{jegou2011pq} with and without exact refinement, and NN-Descent~\cite{dong2011nndescent} for the 10-nearest-neighbor graph. The CPU baselines are DiskANN~\cite{subramanya2019diskann} on the block device, FreshDiskANN~\cite{singh2021freshdiskann}, ACORN~\cite{patel2024acorn}, and Filtered-DiskANN~\cite{golub2024filtereddiskann}. For the LLMs, the baselines are the bf16 and fp16 matrix product with \texttt{torch.topk} of LLM code, exhaustive cuVS search in fp16 and fp32, CAGRA, and IVF-Flat over 4,096 lists. In sparse attention, the strongest baseline is the bf16 matrix product followed by the top-$k$ kernel of FlashInfer~\cite{ye2025flashinfer}. The CertEx engine~\cite{zhao2026certex} serves as the reference for screening without partitions and tiers.

\paragraph{Metrics.}
We report throughput in queries per second at fixed batch sizes, the memory of each tier, and the build and update throughput. For approximate baselines, the exact fraction counts the queries whose $k$ distances equal the exact ones in double precision, which is stricter than Recall@10 under ties. Every answer of \sysname\ is checked against the ground truth and matches it, which gives an exact fraction of 100\% throughout, including on adversarial inputs with exact ties, duplicates, and extreme norms (\supp{supp:stress}).

\subsection{Attention Retrieval on Open LLMs}
\label{sec:eval:llm}

We evaluate exact top-$k$ retrieval over the attention keys of open LLMs, first as an operator on eleven models with one H100 per KV head and then within sparse attention end to end. Long-context inference retrieves these keys for every decoding query~\cite{wu2022memorizing,xiao2024infllm,tang2024quest,liu2025retrievalattention}. LLM code implements the retrieval as a bf16 or fp16 matrix product followed by \texttt{torch.topk}, whose rounding reorders near ties among the largest scores. Figure~\ref{fig:eval_llm} shows that \sysname\ is exact for every query on every model and faster than every exact baseline. On Qwen2.5-7B~\cite{yang2024qwen25}, the approximate CAGRA~\cite{ootomo2024cagra} and IVF-Flat indexes gain throughput only by missing the exact top-10 for many queries, whereas \sysname\ is exact and faster than bf16 and fp16 matrix products (Figure~\ref{fig:eval_llm}(a)). On the eleven models by Qwen~\cite{yang2025qwen3,qwen2026qwen38}, OpenAI~\cite{openai2025gptoss}, Tencent~\cite{tencent2026hy3}, DeepSeek~\cite{deepseek2026v4,deepseek2025v32}, and Zhipu~\cite{glm2026glm5}, \sysname\ answers 135k to 388k queries per second on the 64- to 256-dimensional keys of gpt-oss, Qwen, and Hy3, 5.8$\times$ to 13.1$\times$ exhaustive cuVS~\cite{nvidia2025cuvs} search, and 33k to 45k on the 512- and 576-dimensional keys of DeepSeek and GLM, 2.2$\times$ to 2.8$\times$ the same search (Figure~\ref{fig:eval_llm}(b)). Yet bf16 matrix products are exact for 14\% to 86\% of the queries, fp16 products for 57\% to 98\%, and CAGRA for at most 45\% (Figure~\ref{fig:eval_llm}(c)).

Figure~\ref{fig:eval_llm_k} shows on Qwen2.5-7B~\cite{yang2024qwen25} that \sysname\ stays exact and faster than bf16 matrix products and exhaustive cuVS search up to $k=2{,}048$, which the sparse attention of DeepSeek-V3.2~\cite{deepseek2025v32} uses. At $k=2{,}048$, \sysname\ answers 18.1k exact queries per second, 1.8$\times$ bf16 matrix products and 12.7$\times$ exhaustive cuVS~\cite{nvidia2025cuvs} search in fp32 (Figure~\ref{fig:eval_llm_k}(a)). Only 1.4\% of these queries are exact under bf16 matrix products, against 99.7\% to 99.8\% under cuVS in fp32 (Figure~\ref{fig:eval_llm_k}(b)).

Figure~\ref{fig:eval_llm_e2e} evaluates top-$k$ sparse attention end to end on Qwen2.5-7B-Instruct-1M~\cite{yang2025qwen251m} over the four H100 GPUs. For each of two PG-19~\cite{rae2020pg19} texts, the model prefills one million tokens and evaluates the next 2,048 tokens, whose query heads attend to the attention sinks, to the preceding evaluated tokens, and to the top-$k$ context keys of a selection method. With $k=64$, \sysname\ selects the exact top-64 of every query and shortens the forward pass to 3.6~s, against 11.2~s with bf16 matrix products and the top-$k$ kernel of FlashInfer~\cite{ye2025flashinfer}, 15.9~s with \texttt{torch.topk}, and 7.8~s for dense attention (Figure~\ref{fig:eval_llm_e2e}(a)). Both bf16 selections pick the exact top-64 for only 42\% to 43\% of the queries (Figure~\ref{fig:eval_llm_e2e}(b)). With $k=256$, the forward pass takes 7.6~s against 12.1~s and 16.8~s. The perplexities of the four selections lie within 0.3\% of each other on both texts. \sysname\ thus makes the selection exact at no cost in quality and in less time. The exact selection also makes the attended keys a function of the model alone, whereas bf16 and fp16 matrix products attend to different keys for the same query.

\begin{figure}[t]
\centering
\includegraphics[width=\columnwidth]{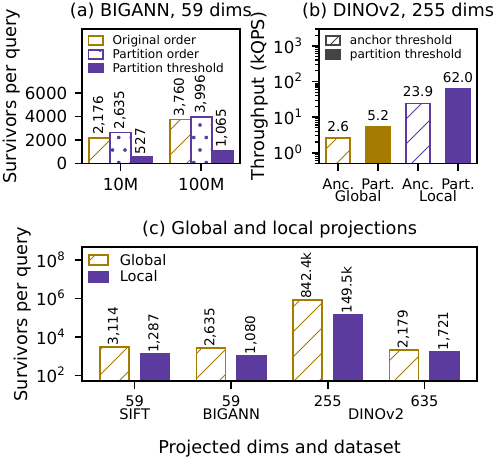}
\caption{Partition thresholds and local projections.}
\label{fig:eval_bounds}
\end{figure}

\subsection{Tighter Bounds from Partitions}
\label{sec:eval:ivf_pruning}
\label{sec:eval:local_bounds}

\begin{figure}[t]
\centering
\includegraphics[width=\columnwidth]{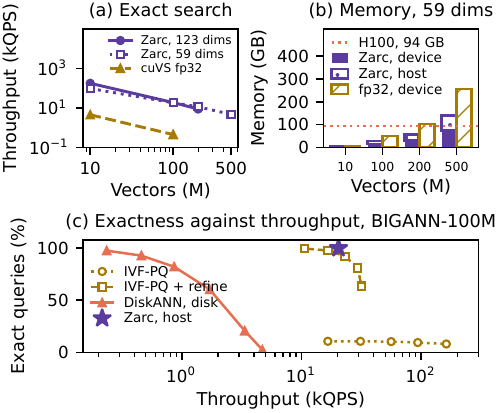}
\caption{Tiered memory on BIGANN with one H100.}
\label{fig:eval_tiered}
\end{figure}

We evaluate partition thresholds and local projections on BIGANN~\cite{jegou2011bigann}, where screening dominates the batch time, and on DINOv2~\cite{oquab2024dinov2}, where verification dominates it. Both mechanisms change only the survivors, which keeps every variant exact.

Figure~\ref{fig:eval_bounds} shows that the two mechanisms cut the survivors and raise the throughput where survivors are expensive. On BIGANN~\cite{jegou2011bigann} with 59-dimensional projections, the partition threshold leaves 5.0$\times$ and 3.8$\times$ fewer survivors at 10M and 100M (Figure~\ref{fig:eval_bounds}(a)), although the throughput changes little there, as screening dominates. On DINOv2~\cite{oquab2024dinov2} with 255-dimensional projections, local projections and the partition threshold together raise the throughput 24$\times$, from 2.6k to 62.0k queries per second (Figure~\ref{fig:eval_bounds}(b)). Local projections alone cut the survivors per query from 842k to 150k on DINOv2 and by 2.4$\times$ on SIFT1M and BIGANN-10M (Figure~\ref{fig:eval_bounds}(c)).

\subsection{Tiered Memory Scaling}
\label{sec:eval:tiered}

We evaluate the placement of the raw vectors in host memory on BIGANN~\cite{jegou2011bigann} prefixes from 10M to 500M vectors on one H100, with the screening rows and an anchor sample of every 16th vector in device memory. Exhaustive fp32 search on the same GPU stops at about 190M vectors.

Figure~\ref{fig:eval_tiered} shows that host memory extends exact search to 500 million vectors on one GPU. At 10M and 100M, where exhaustive fp32 search still fits on the GPU, \sysname\ runs 39$\times$ and 42$\times$ faster than it and keeps answering beyond that size, with 5,233 queries per second at 500M vectors and 59-dimensional projections (Figure~\ref{fig:eval_tiered}(a)). Each query reads only 699 of its 5,593 survivors (\supp{supp:ordered}). At 500M vectors, \sysname\ needs 71~GB of device memory, against the 256~GB that exhaustive fp32 search would need (Figure~\ref{fig:eval_tiered}(b)). Compressed and disk-resident indexes reach this capacity only by giving up exactness (Figure~\ref{fig:eval_tiered}(c)). cuVS~\cite{nvidia2025cuvs} IVF-PQ~\cite{jegou2011pq} needs exact refinement with 512 probes to be exact for 99.6\% of the queries at 10.6k queries per second, against 20.3k for \sysname. DiskANN~\cite{subramanya2019diskann} on the block device, in turn, is exact for at most 97.6\% of the queries at 236 queries per second.

\subsection{Multi-GPU Scalability}
\label{sec:eval:multigpu}

We evaluate the sharded search on the four H100 GPUs of the main server, with the raw vectors of every shard in pinned host memory and exhaustive cuVS~\cite{nvidia2025cuvs} search over half-precision shards as the baseline. The shared threshold keeps every answer exact on every shard count.

\begin{figure}[t]
\centering
\includegraphics[width=\columnwidth]{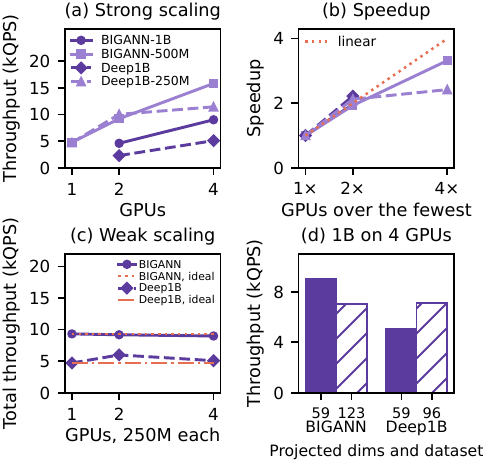}
\caption{Sharded exact search on four H100 GPUs.}
\label{fig:eval_multigpu}
\end{figure}

Figure~\ref{fig:eval_multigpu} shows that the shared threshold scales exact search to one billion vectors. On BIGANN-500M~\cite{jegou2011bigann}, \sysname\ answers 4,759, 9,188, and 15,796 queries per second on one, two, and four GPUs (Figure~\ref{fig:eval_multigpu}(a)). On BIGANN-1B, four GPUs reach 1.96$\times$ the throughput of the two GPUs that first hold it (Figure~\ref{fig:eval_multigpu}(b)), with 8,997 queries per second, 269$\times$ exhaustive cuVS~\cite{nvidia2025cuvs} search (Figure~\ref{fig:opener}(a)). Under weak scaling at 250M vectors per GPU, the total throughput of all GPUs stays within 3.6\% up to one billion vectors (Figure~\ref{fig:eval_multigpu}(c)). This flat line is the ideal, since every query screens a dataset that grows with the GPUs. The best number of projected dimensions depends on the dataset (Figure~\ref{fig:eval_multigpu}(d)). On BIGANN-1B, 59 dimensions answer 8,997 queries per second against 7,046 for 123 dimensions, since the wider rows double the screened bytes. On Deep1B~\cite{babenko2016deep}, the full 96 dimensions answer 7,133 against 5,087 for 59 dimensions, as they remove survivors that verification would otherwise fetch from host memory as 384-byte vectors.

\begin{figure}[t]
\centering
\includegraphics[width=\columnwidth]{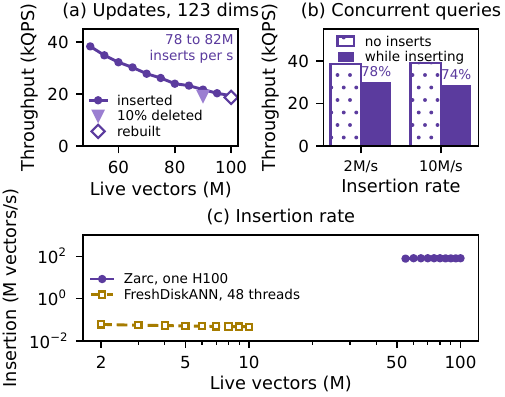}
\caption{Updates and concurrent queries on BIGANN.}
\label{fig:eval_updates}
\end{figure}

\subsection{Streaming Dynamic Updates}
\label{sec:eval:updates}

\begin{figure}[t]
\centering
\includegraphics[width=\columnwidth]{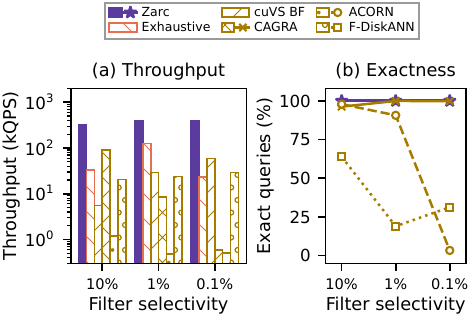}
\caption{Exact filtered search on BIGANN-10M.}
\label{fig:eval_filtered}
\end{figure}

We evaluate updates by growing a BIGANN~\cite{jegou2011bigann} index from 50M to 100M vectors in ten insertion steps and then deleting 10\% of the vectors, with every query checked against exhaustive search over the live vectors. FreshDiskANN~\cite{singh2021freshdiskann} on 48 CPU threads serves as the baseline for insertions into a graph index.

Figure~\ref{fig:eval_updates} shows that updates keep every answer exact at the throughput of a rebuilt index. While insertions run at 78 to 82 million vectors per second, the throughput follows the dataset size from 38.3k queries per second at 50M to 19.4k at 100M, which matches the 18.7k of an index rebuilt at 100M (Figure~\ref{fig:eval_updates}(a)). While a background thread inserts 25 million vectors at 2 or 10 million vectors per second, the queries keep 78\% and 74\% of their throughput, with every one of the 170,000 checked answers exact (Figure~\ref{fig:eval_updates}(b)). FreshDiskANN~\cite{singh2021freshdiskann} on 48 CPU threads inserts about 1,300$\times$ slower (Figure~\ref{fig:eval_updates}(c)), as it searches the graph and repairs the edges around every new vector.

\subsection{Application Workloads}
\label{sec:eval:apps}
\label{sec:eval:filtered}
\label{sec:eval:rag}

\paragraph{Filtered search.}
Figure~\ref{fig:eval_filtered} shows that \sysname\ is the fastest exact method at every selectivity on BIGANN-10M~\cite{jegou2011bigann}, with 324k to 403k queries per second, 6.8$\times$ to 58$\times$ exhaustive cuVS~\cite{nvidia2025cuvs} search over the matching vectors (Figure~\ref{fig:eval_filtered}(a)). At its most accurate setting, CAGRA~\cite{ootomo2024cagra} runs 46$\times$ and 692$\times$ slower than \sysname\ at 1\% and 0.1\% selectivity, because every label requires a separate CAGRA call. In contrast, the filtered graph indexes ACORN~\cite{patel2024acorn} and Filtered-DiskANN~\cite{golub2024filtereddiskann} return the exact neighbors of at most 97.9\% and 63.9\% of the queries (Figure~\ref{fig:eval_filtered}(b)).

\paragraph{Semantic deduplication.}
\sysname\ builds the exact 10-NN graph of DataComp-12.8M~\cite{gadre2023datacomp} in 8.2 minutes end to end (Figure~\ref{fig:opener}(c) and \supp{supp:allknn}), 42$\times$ faster than the projected 5.8 hours of exhaustive cuVS~\cite{nvidia2025cuvs} search and 58$\times$ faster than the CertEx engine~\cite{zhao2026certex}. The approximate cuVS NN-Descent~\cite{dong2011nndescent} leaves at least 3.0\% of the rows inexact, unflagged. The partition threshold cuts survivors per query from 10,486 to 114, and verification reads 41.

\begin{figure}[t]
\centering
\includegraphics[width=\columnwidth]{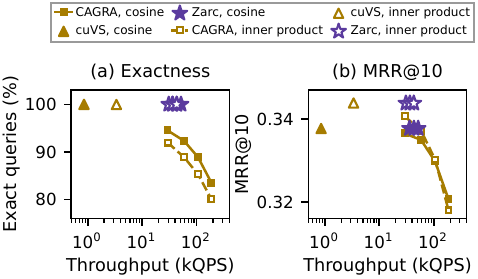}
\caption{Exact retrieval on MS MARCO.}
\label{fig:eval_rag}
\end{figure}

\paragraph{Passage retrieval.}
Retrieval-augmented generation (RAG)~\cite{lewis2020rag} tolerates approximate neighbors. Consequently, exact search must match the speed of approximate indexes there. Figure~\ref{fig:eval_rag} shows that \sysname\ is exact on the 8.84 million passages of MS MARCO~\cite{nguyen2016msmarco} under both cosine similarity and inner products and faster than CAGRA at its most accurate setting. Under cosine similarity, \sysname\ answers 35.9k to 51.7k queries per second, 60$\times$ exhaustive cuVS~\cite{nvidia2025cuvs} search, whereas CAGRA~\cite{ootomo2024cagra} at its most accurate setting misses 5.5\% of the queries at 29.2k. Under inner products, \sysname\ answers 30.8k to 43.0k, 13$\times$ exhaustive cuVS search, while CAGRA misses 8.0\% of the queries (Figure~\ref{fig:eval_rag}(a)). Exactness carries over to retrieval quality, which MS MARCO measures by the mean reciprocal rank at 10 (MRR@10). MRR@10 averages over the queries the reciprocal rank of the first relevant passage among the top 10 results. Under cosine similarity, \sysname\ reaches the MRR@10 of exhaustive search, 0.338, against 0.321 to 0.337 for CAGRA (Figure~\ref{fig:eval_rag}(b)).

\section{Related Work}
\label{sec:related_work}
\label{sec:rel:exact}
\label{sec:rel:gpu_disk}
\label{sec:rel:filtered_certex}

\paragraph{Exact search.}
Metric access methods such as AESA~\cite{vidal1986aesa}, cover trees~\cite{beygelzimer2006cover}, VP-trees~\cite{yianilos1993vptree}, M-trees~\cite{ciaccia1997mtree}, and FLANN~\cite{muja2014flann} and bound-based filters such as GEMINI~\cite{faloutsos1994gemini}, the VA-file~\cite{weber1998vafile}, and iDistance~\cite{jagadish2005idistance} prune candidates by pointer chasing and scalar comparisons, which leave the matrix units of a GPU idle. Exhaustive GPU search, from \citet{garcia2008gpuknn} to Faiss~\cite{johnson2021faiss} and cuVS~\cite{nvidia2025cuvs}, uses these matrix units but evaluates every distance in single precision, which can misrank near ties. Floating-point filters in computational geometry~\cite{fortune1993exact,shewchuk1997adaptive} and validated numerics~\cite{higham2002accuracy} resolve such cases cheaply by evaluating in fast floating point and resorting to exact arithmetic only where the error bound requires it. \sysname\ applies such rounding-error bounds to the half-precision tensor-core products that screen every vector.

\paragraph{Approximate and disk-resident indexes.}
Inverted files~\cite{johnson2021faiss,babenko2012imi,baranchuk2018revisiting} over vectors compressed by OPQ~\cite{ge2013opq}, ScaNN~\cite{guo2020scann}, or RaBitQ~\cite{gao2024rabitq} and proximity graphs such as HNSW~\cite{malkov2020hnsw}, NSG~\cite{fu2019nsg}, CAGRA~\cite{ootomo2024cagra}, SONG~\cite{zhao2020song}, and GGNN~\cite{groh2023ggnn} trade recall for throughput through the probe count or the search width. HM-ANN~\cite{ren2020hmann}, BANG~\cite{venkatasubba2025bang}, and DeepStore~\cite{lee2024deepstore} place billion-scale indexes in heterogeneous memory, on one GPU, or in storage, which BaM~\cite{gao2023bam} makes directly accessible to GPUs. DiskANN~\cite{subramanya2019diskann}, Starling~\cite{wang2024starling}, SPANN~\cite{chen2021spann}, and FreshDiskANN~\cite{singh2021freshdiskann} keep the full vectors on solid-state storage, whose reads Rummy~\cite{zhang2024rummy} and FusionANNS~\cite{tian2025fusionanns} overlap with distance computation on the CPU and the GPU. PilotANN~\cite{gui2026pilotann}, SVFusion~\cite{peng2026svfusion}, HAVEN~\cite{hsu2026haven}, and IVF-RaBitQ~\cite{shi2026ivfrabitq} spread approximate search over the memory hierarchy of a GPU server, where the tier of a vector decides its accuracy, whereas the tiers of \sysname\ decide only where a vector is read from.

\paragraph{Filtered search.}
Vector databases such as Milvus~\cite{wang2021milvus}, Manu~\cite{guo2022manu}, AnalyticDB-V~\cite{wei2020analyticdbv}, and VBase~\cite{zhang2023vbase} integrate relational predicates with vector indexes. At the index level, ACORN~\cite{patel2024acorn} and Filtered-DiskANN~\cite{golub2024filtereddiskann} add predicate-agnostic or label-aware edges to graph indexes. Both kinds of systems depend on the connectivity of the matching vertices. \sysname, in contrast, screens only the rows of the queried label.

\paragraph{Certified exclusion.}
The error bounds of RaBitQ~\cite{gao2024rabitq,gao2025extrabitq} and the conformal recall guarantees of ConANN~\cite{horchidan2025conann,vovk2005conformal} hold over a distribution of rotations or queries, whereas those of \sysname\ hold for every query and every vector. These bounds come from CertEx~\cite{zhao2026certex}, which proposed certified exclusion for data that fit on one GPU. \sysname\ extends them with partition thresholds, memory tiers, sharding, concurrent updates, filters, and inner products.

\section{Conclusion}
\label{sec:conclusion}

This paper presented \sysname, which makes exact vector search practical at the scale of billion-vector datasets and million-token contexts. Its central observation is that certified exclusion needs only the short half-precision screening rows in GPU memory, whereas the few surviving vectors can stay in host memory. On billion-scale benchmarks, eleven open LLMs, and sparse attention over one million tokens, \sysname\ returns the exact answer for every query while it runs up to 270$\times$ faster than exhaustive GPU search and, in retrieval, faster than approximate indexes at their most accurate settings. Since \sysname\ keeps the calls of cuVS, Faiss, PyTorch, and FlashInfer, adopting it changes little more than an import.

\bibliography{refs}

\begin{thebibliography}{92}
\providecommand{\natexlab}[1]{#1}
\providecommand{\url}[1]{\texttt{#1}}
\expandafter\ifx\csname urlstyle\endcsname\relax
  \providecommand{\doi}[1]{doi: #1}\else
  \providecommand{\doi}{doi: \begingroup \urlstyle{rm}\Url}\fi

\bibitem[Abbas et~al.(2023)Abbas, Tirumala, Simig, Ganguli, and
  Morcos]{abbas2023semdedup}
Abbas, A., Tirumala, K., Simig, D., Ganguli, S., and Morcos, A.~S.
\newblock {SemDeDup: Data-efficient learning at web-scale through semantic
  deduplication}.
\newblock \emph{arXiv preprint arXiv:2303.09540}, 2023.
\newblock URL \url{https://arxiv.org/abs/2303.09540}.

\bibitem[Adinets \& Merrill(2022)Adinets and Merrill]{adinets2022onesweep}
Adinets, A. and Merrill, D.
\newblock Onesweep: A faster least significant digit radix sort for {GPUs}.
\newblock \emph{arXiv preprint arXiv:2206.01784}, 2022.
\newblock \doi{10.48550/arXiv.2206.01784}.
\newblock URL \url{https://arxiv.org/abs/2206.01784}.

\bibitem[Aumüller et~al.(2020)Aumüller, Bernhardsson, and
  Faithfull]{aumuller2020annbench}
Aumüller, M., Bernhardsson, E., and Faithfull, A.
\newblock {ANN-Benchmarks: A benchmarking tool for approximate nearest neighbor
  algorithms}.
\newblock \emph{Information Systems}, 87:\penalty0 101374, 2020.
\newblock URL \url{https://doi.org/10.1016/j.is.2019.02.006}.

\bibitem[Babenko \& Lempitsky(2012)Babenko and Lempitsky]{babenko2012imi}
Babenko, A. and Lempitsky, V.
\newblock {The inverted multi-index}.
\newblock In \emph{2012 IEEE Conference on Computer Vision and Pattern
  Recognition}, pp.\  3069--3076, 2012.
\newblock URL \url{https://doi.org/10.1109/CVPR.2012.6248038}.

\bibitem[Babenko \& Lempitsky(2016)Babenko and Lempitsky]{babenko2016deep}
Babenko, A. and Lempitsky, V.
\newblock Efficient indexing of billion-scale datasets of deep descriptors.
\newblock In \emph{Proceedings of the IEEE Conference on Computer Vision and
  Pattern Recognition (CVPR)}, pp.\  2055--2063, 2016.
\newblock URL \url{https://doi.org/10.1109/CVPR.2016.226}.

\bibitem[Baranchuk et~al.(2018)Baranchuk, Babenko, and
  Malkov]{baranchuk2018revisiting}
Baranchuk, D., Babenko, A., and Malkov, Y.
\newblock Revisiting the inverted indices for billion-scale approximate nearest
  neighbors.
\newblock In \emph{Proceedings of the European Conference on Computer Vision
  (ECCV)}, pp.\  209--224, 2018.
\newblock URL \url{https://doi.org/10.1007/978-3-030-01258-8_13}.

\bibitem[Beygelzimer et~al.(2006)Beygelzimer, Kakade, and
  Langford]{beygelzimer2006cover}
Beygelzimer, A., Kakade, S., and Langford, J.
\newblock {Cover trees for nearest neighbor}.
\newblock In \emph{Proceedings of the 23rd international conference on Machine
  learning}, pp.\  97--104, 2006.
\newblock URL \url{https://doi.org/10.1145/1143844.1143857}.

\bibitem[Blanchard et~al.(2020)Blanchard, Higham, Lopez, Mary, and
  Pranesh]{blanchard2020tensorcore}
Blanchard, P., Higham, N.~J., Lopez, F., Mary, T., and Pranesh, S.
\newblock Mixed precision block fused multiply-add: Error analysis and
  application to {GPU} tensor cores.
\newblock \emph{SIAM Journal on Scientific Computing}, 42\penalty0
  (3):\penalty0 C124--C141, 2020.
\newblock URL \url{https://doi.org/10.1137/19M1289546}.

\bibitem[Chen et~al.(2021)Chen, Zhao, Wang, Li, Liu, Li, Yang, and
  Wang]{chen2021spann}
Chen, Q., Zhao, B., Wang, H., Li, M., Liu, C., Li, Z., Yang, M., and Wang, J.
\newblock {SPANN}: Highly-efficient {Billion-scale} approximate nearest
  neighbor search.
\newblock In \emph{Advances in Neural Information Processing Systems
  (NeurIPS)}, volume~34, pp.\  5199--5212, 2021.
\newblock URL
  \url{https://proceedings.neurips.cc/paper/2021/hash/299dc35e747eb77177d9cea10a802da2-Abstract.html}.

\bibitem[Ciaccia et~al.(1997)Ciaccia, Patella, and Zezula]{ciaccia1997mtree}
Ciaccia, P., Patella, M., and Zezula, P.
\newblock {M-tree}: An efficient access method for similarity search in metric
  spaces.
\newblock In \emph{Proceedings of the 23rd International Conference on Very
  Large Data Bases (VLDB)}, pp.\  426--435, 1997.
\newblock URL \url{http://www.vldb.org/conf/1997/P426.PDF}.

\bibitem[{DeepSeek-AI}(2024)]{deepseek2024v2}
{DeepSeek-AI}.
\newblock {DeepSeek-V2}: A strong, economical, and efficient mixture-of-experts
  language model.
\newblock \emph{arXiv preprint arXiv:2405.04434}, 2024.
\newblock URL \url{https://arxiv.org/abs/2405.04434}.

\bibitem[{DeepSeek-AI}(2025)]{deepseek2025v32}
{DeepSeek-AI}.
\newblock {DeepSeek-V3.2}: Pushing the frontier of open large language models.
\newblock \emph{arXiv preprint arXiv:2512.02556}, 2025.
\newblock URL \url{https://arxiv.org/abs/2512.02556}.

\bibitem[{DeepSeek-AI}(2026)]{deepseek2026v4}
{DeepSeek-AI}.
\newblock {DeepSeek-V4}: Towards highly efficient million-token context
  intelligence.
\newblock \emph{arXiv preprint arXiv:2606.19348}, 2026.
\newblock URL \url{https://arxiv.org/abs/2606.19348}.

\bibitem[Deng et~al.(2009)Deng, Dong, Socher, Li, Li, and
  Fei-Fei]{deng2009imagenet}
Deng, J., Dong, W., Socher, R., Li, L.-J., Li, K., and Fei-Fei, L.
\newblock {ImageNet}: A large-scale hierarchical image database.
\newblock In \emph{IEEE Conference on Computer Vision and Pattern Recognition
  (CVPR)}, pp.\  248--255, 2009.
\newblock URL \url{https://doi.org/10.1109/CVPR.2009.5206848}.

\bibitem[Dong et~al.(2011)Dong, Charikar, and Li]{dong2011nndescent}
Dong, W., Charikar, M., and Li, K.
\newblock Efficient $k$-nearest neighbor graph construction for generic
  similarity measures.
\newblock In \emph{Proceedings of the 20th International Conference on World
  Wide Web (WWW)}, pp.\  577--586, 2011.
\newblock URL \url{https://doi.org/10.1145/1963405.1963487}.

\bibitem[Douze et~al.(2026)Douze, Guzhva, Deng, Johnson, Szilvasy, Mazar{\'e},
  Lomeli, Hosseini, and J{\'e}gou]{douze2024faisslib}
Douze, M., Guzhva, A., Deng, C., Johnson, J., Szilvasy, G., Mazar{\'e}, P.-E.,
  Lomeli, M., Hosseini, L., and J{\'e}gou, H.
\newblock The {Faiss} library.
\newblock \emph{IEEE Transactions on Big Data}, 12\penalty0 (2):\penalty0
  346--361, 2026.
\newblock URL \url{https://doi.org/10.1109/TBDATA.2025.3618474}.

\bibitem[Faloutsos et~al.(1994)Faloutsos, Ranganathan, and
  Manolopoulos]{faloutsos1994gemini}
Faloutsos, C., Ranganathan, M., and Manolopoulos, Y.
\newblock {Fast subsequence matching in time-series databases}.
\newblock In \emph{Proceedings of the 1994 ACM SIGMOD international conference
  on Management of data}, pp.\  419--429, 1994.
\newblock URL \url{https://doi.org/10.1145/191839.191925}.

\bibitem[Fortune \& Van~Wyk(1993)Fortune and Van~Wyk]{fortune1993exact}
Fortune, S. and Van~Wyk, C.~J.
\newblock Efficient exact arithmetic for computational geometry.
\newblock In \emph{Proceedings of the Ninth Annual Symposium on Computational
  Geometry (SoCG)}, pp.\  163--172, 1993.
\newblock URL \url{https://doi.org/10.1145/160985.161015}.

\bibitem[Fu et~al.(2019)Fu, Xiang, Wang, and Cai]{fu2019nsg}
Fu, C., Xiang, C., Wang, C., and Cai, D.
\newblock Fast approximate nearest neighbor search with the navigating
  spreading-out graph.
\newblock \emph{Proceedings of the VLDB Endowment}, 12\penalty0 (5):\penalty0
  461--474, 2019.
\newblock URL \url{https://doi.org/10.14778/3303753.3303754}.

\bibitem[Gadre et~al.(2023)Gadre, Ilharco, Fang, Hayase, Smyrnis, Nguyen,
  Marten, Wortsman, Ghosh, Zhang, Orgad, Entezari, Daras, Pratt, Ramanujan,
  Bitton, Marathe, Mussmann, Vencu, Cherti, Krishna, Koh, Saukh, Ratner, Song,
  Hajishirzi, Farhadi, Beaumont, Oh, Dimakis, Jitsev, Carmon, Shankar, and
  Schmidt]{gadre2023datacomp}
Gadre, S.~Y., Ilharco, G., Fang, A., Hayase, J., Smyrnis, G., Nguyen, T.,
  Marten, R., Wortsman, M., Ghosh, D., Zhang, J., Orgad, E., Entezari, R.,
  Daras, G., Pratt, S., Ramanujan, V., Bitton, Y., Marathe, K., Mussmann, S.,
  Vencu, R., Cherti, M., Krishna, R., Koh, P.~W., Saukh, O., Ratner, A., Song,
  S., Hajishirzi, H., Farhadi, A., Beaumont, R., Oh, S., Dimakis, A., Jitsev,
  J., Carmon, Y., Shankar, V., and Schmidt, L.
\newblock {DataComp: In search of the next generation of multimodal datasets}.
\newblock In \emph{Advances in Neural Information Processing Systems
  (NeurIPS)}, 2023.
\newblock URL \url{https://doi.org/10.52202/075280-1179}.

\bibitem[Gao \& Long(2024)Gao and Long]{gao2024rabitq}
Gao, J. and Long, C.
\newblock {RaBitQ: Quantizing High-Dimensional Vectors with a Theoretical Error
  Bound for Approximate Nearest Neighbor Search}.
\newblock \emph{Proceedings of the ACM on Management of Data}, 2\penalty0
  (3):\penalty0 1--27, 2024.
\newblock URL \url{https://doi.org/10.1145/3654970}.

\bibitem[Gao et~al.(2025)Gao, Gou, Xu, Yang, Long, and Wong]{gao2025extrabitq}
Gao, J., Gou, Y., Xu, Y., Yang, Y., Long, C., and Wong, R. C.-W.
\newblock {Practical and Asymptotically Optimal Quantization of
  High-Dimensional Vectors in Euclidean Space for Approximate Nearest Neighbor
  Search}.
\newblock \emph{Proceedings of the ACM on Management of Data}, 2025.
\newblock URL \url{https://doi.org/10.1145/3725413}.

\bibitem[Garcia et~al.(2008)Garcia, Debreuve, and Barlaud]{garcia2008gpuknn}
Garcia, V., Debreuve, E., and Barlaud, M.
\newblock {Fast k nearest neighbor search using GPU}.
\newblock In \emph{2008 IEEE Computer Society Conference on Computer Vision and
  Pattern Recognition Workshops}, pp.\  1--6, 2008.
\newblock URL \url{https://doi.org/10.1109/cvprw.2008.4563100}.

\bibitem[Ge et~al.(2013)Ge, He, Ke, and Sun]{ge2013opq}
Ge, T., He, K., Ke, Q., and Sun, J.
\newblock {Optimized Product Quantization for Approximate Nearest Neighbor
  Search}.
\newblock In \emph{IEEE Conference on Computer Vision and Pattern Recognition
  (CVPR)}, pp.\  2946--2953, 2013.
\newblock URL \url{https://doi.org/10.1109/CVPR.2013.379}.

\bibitem[{GLM-5 Team}(2026)]{glm2026glm5}
{GLM-5 Team}.
\newblock {GLM-5}: from vibe coding to agentic engineering.
\newblock \emph{arXiv preprint arXiv:2602.15763}, 2026.
\newblock URL \url{https://arxiv.org/abs/2602.15763}.

\bibitem[Gollapudi et~al.(2023)Gollapudi, Karia, Sivashankar, Krishnaswamy,
  Begwani, Raz, Lin, Zhang, Mahapatro, Srinivasan, Singh, and
  Simhadri]{golub2024filtereddiskann}
Gollapudi, S., Karia, N., Sivashankar, V., Krishnaswamy, R., Begwani, N., Raz,
  S., Lin, Y., Zhang, Y., Mahapatro, N., Srinivasan, P., Singh, A., and
  Simhadri, H.~V.
\newblock {Filtered-DiskANN}: Graph algorithms for approximate nearest neighbor
  search with filters.
\newblock In \emph{Proceedings of the ACM Web Conference 2023 (WWW)}, pp.\
  3406--3416, 2023.
\newblock URL \url{https://doi.org/10.1145/3543507.3583552}.

\bibitem[Groh et~al.(2023)Groh, Ruppert, Wieschollek, and Lensch]{groh2023ggnn}
Groh, F., Ruppert, L., Wieschollek, P., and Lensch, H. P.~A.
\newblock {GGNN: Graph-Based GPU Nearest Neighbor Search}.
\newblock \emph{IEEE Transactions on Big Data}, 9\penalty0 (1):\penalty0
  267--279, 2023.
\newblock URL \url{https://doi.org/10.1109/tbdata.2022.3161156}.

\bibitem[Gui et~al.(2026)Gui, Yin, Yan, Zhang, Zhang, and
  Cheng]{gui2026pilotann}
Gui, Y., Yin, P., Yan, X., Zhang, C., Zhang, W., and Cheng, J.
\newblock Pilotann: Memory-bounded gpu acceleration for vector search.
\newblock In \emph{Proceedings of the 32nd ACM SIGKDD Conference on Knowledge
  Discovery and Data Mining V.1}, KDD '26, pp.\  348–358. Association for
  Computing Machinery, 2026.
\newblock URL \url{https://doi.org/10.1145/3770854.3780192}.

\bibitem[Guo et~al.(2020)Guo, Sun, Lindgren, Geng, Simcha, Chern, and
  Kumar]{guo2020scann}
Guo, R., Sun, P., Lindgren, E., Geng, Q., Simcha, D., Chern, F., and Kumar, S.
\newblock {Accelerating Large-Scale Inference with Anisotropic Vector
  Quantization}.
\newblock In \emph{International Conference on Machine Learning (ICML)}, volume
  119 of \emph{Proceedings of Machine Learning Research}, pp.\  3887--3896,
  2020.
\newblock URL \url{https://proceedings.mlr.press/v119/guo20h.html}.

\bibitem[Guo et~al.(2022)Guo, Luan, Xiang, Yan, Yi, Luo, Cheng, Xu, Luo, Liu,
  Cao, Qiao, Wang, Tang, and Xie]{guo2022manu}
Guo, R., Luan, X., Xiang, L., Yan, X., Yi, X., Luo, J., Cheng, Q., Xu, W., Luo,
  J., Liu, F., Cao, Z., Qiao, Y., Wang, T., Tang, B., and Xie, C.
\newblock Manu: A cloud native vector database management system.
\newblock \emph{Proceedings of the VLDB Endowment}, 15\penalty0 (12):\penalty0
  3548--3561, 2022.
\newblock URL \url{https://doi.org/10.14778/3554821.3554843}.

\bibitem[Higham(2002)]{higham2002accuracy}
Higham, N.~J.
\newblock \emph{Accuracy and Stability of Numerical Algorithms}.
\newblock Society for Industrial and Applied Mathematics (SIAM), Philadelphia,
  PA, second edition, 2002.
\newblock \doi{10.1137/1.9780898718027}.
\newblock URL \url{https://doi.org/10.1137/1.9780898718027}.

\bibitem[Hofst{\"a}tter et~al.(2021)Hofst{\"a}tter, Lin, Yang, Lin, and
  Hanbury]{hofstatter2021tasb}
Hofst{\"a}tter, S., Lin, S.-C., Yang, J.-H., Lin, J., and Hanbury, A.
\newblock {Efficiently Teaching an Effective Dense Retriever with Balanced
  Topic Aware Sampling}.
\newblock In \emph{Proceedings of the 44th International ACM SIGIR Conference
  on Research and Development in Information Retrieval}, pp.\  113--122, 2021.
\newblock URL \url{https://doi.org/10.1145/3404835.3462891}.

\bibitem[Horchidan et~al.(2025)Horchidan, Zeiher, Bostr{\"o}m, and
  Carbone]{horchidan2025conann}
Horchidan, S., Zeiher, F., Bostr{\"o}m, H., and Carbone, P.
\newblock {ConANN}: Conformal approximate nearest neighbor search.
\newblock \emph{Proceedings of the VLDB Endowment}, 19\penalty0 (1):\penalty0
  29--42, 2025.
\newblock URL \url{https://www.vldb.org/pvldb/vol19/p29-horchidan.pdf}.

\bibitem[Hsu et~al.(2026)Hsu, Xu, Liu, Rosing, and Yu]{hsu2026haven}
Hsu, P.-K., Xu, W., Liu, Q., Rosing, T., and Yu, S.
\newblock {HAVEN}: High-bandwidth flash augmented vector engine for large-scale
  approximate nearest-neighbor search acceleration.
\newblock \emph{arXiv preprint arXiv:2603.01175}, 2026.
\newblock URL \url{https://arxiv.org/abs/2603.01175}.

\bibitem[Jagadish et~al.(2005)Jagadish, Ooi, Tan, Yu, and
  Zhang]{jagadish2005idistance}
Jagadish, H.~V., Ooi, B.~C., Tan, K.-L., Yu, C., and Zhang, R.
\newblock {iDistance}: An adaptive b+-tree based indexing method for nearest
  neighbor search.
\newblock \emph{ACM Transactions on Database Systems (TODS)}, 30\penalty0
  (2):\penalty0 364--397, 2005.
\newblock URL \url{https://doi.org/10.1145/1071610.1071612}.

\bibitem[Jegou et~al.(2011)Jegou, Tavenard, Douze, and
  Amsaleg]{jegou2011bigann}
Jegou, H., Tavenard, R., Douze, M., and Amsaleg, L.
\newblock {Searching in one billion vectors: Re-rank with source coding}.
\newblock In \emph{2011 IEEE International Conference on Acoustics, Speech and
  Signal Processing (ICASSP)}, pp.\  861--864, 2011.
\newblock URL \url{https://doi.org/10.1109/ICASSP.2011.5946540}.

\bibitem[Johnson et~al.(2021)Johnson, Douze, and Jegou]{johnson2021faiss}
Johnson, J., Douze, M., and Jegou, H.
\newblock {Billion-Scale Similarity Search with GPUs}.
\newblock \emph{IEEE Transactions on Big Data}, 7\penalty0 (3):\penalty0
  535--547, 2021.
\newblock URL \url{https://doi.org/10.1109/TBDATA.2019.2921572}.

\bibitem[Jégou et~al.(2011)Jégou, Douze, and Schmid]{jegou2011pq}
Jégou, H., Douze, M., and Schmid, C.
\newblock {Product Quantization for Nearest Neighbor Search}.
\newblock \emph{IEEE Transactions on Pattern Analysis and Machine
  Intelligence}, 33\penalty0 (1):\penalty0 117--128, 2011.
\newblock URL \url{https://doi.org/10.1109/TPAMI.2010.57}.

\bibitem[Khandelwal et~al.(2020)Khandelwal, Levy, Jurafsky, Zettlemoyer, and
  Lewis]{khandelwal2020knnlm}
Khandelwal, U., Levy, O., Jurafsky, D., Zettlemoyer, L., and Lewis, M.
\newblock Generalization through memorization: Nearest neighbor language
  models.
\newblock In \emph{International Conference on Learning Representations
  (ICLR)}, 2020.
\newblock URL \url{https://openreview.net/forum?id=HklBjCEKvH}.

\bibitem[Kwon et~al.(2023)Kwon, Li, Zhuang, Sheng, Zheng, Yu, Gonzalez, Zhang,
  and Stoica]{kwon2023vllm}
Kwon, W., Li, Z., Zhuang, S., Sheng, Y., Zheng, L., Yu, C.~H., Gonzalez, J.~E.,
  Zhang, H., and Stoica, I.
\newblock Efficient memory management for large language model serving with
  {PagedAttention}.
\newblock In \emph{Proceedings of the 29th Symposium on Operating Systems
  Principles (SOSP)}, pp.\  611--626, 2023.
\newblock URL \url{https://doi.org/10.1145/3600006.3613165}.

\bibitem[Lewis et~al.(2020)Lewis, Perez, Piktus, Petroni, Karpukhin, Goyal,
  K{\"u}ttler, Lewis, Yih, Rockt{\"a}schel, Riedel, and Kiela]{lewis2020rag}
Lewis, P., Perez, E., Piktus, A., Petroni, F., Karpukhin, V., Goyal, N.,
  K{\"u}ttler, H., Lewis, M., Yih, W., Rockt{\"a}schel, T., Riedel, S., and
  Kiela, D.
\newblock Retrieval-augmented generation for knowledge-intensive {NLP} tasks.
\newblock In \emph{Advances in Neural Information Processing Systems
  (NeurIPS)}, volume~33, pp.\  9459--9474, 2020.
\newblock URL
  \url{https://proceedings.neurips.cc/paper/2020/hash/6b493230205f780e1bc26945df7481e5-Abstract.html}.

\bibitem[Liu et~al.(2025)Liu, Chen, Lu, Jiang, Han, Zhang, Chen, Zhang, Ding,
  Zhang, Chen, Yang, Yang, and Qiu]{liu2025retrievalattention}
Liu, D., Chen, M., Lu, B., Jiang, H., Han, Z., Zhang, Q., Chen, Q., Zhang, C.,
  Ding, B., Zhang, K., Chen, C., Yang, F., Yang, Y., and Qiu, L.
\newblock {RetrievalAttention}: Accelerating long-context {LLM} inference via
  vector retrieval.
\newblock In \emph{Advances in Neural Information Processing Systems
  (NeurIPS)}, 2025.
\newblock URL
  \url{https://papers.nips.cc/paper_files/paper/2025/hash/4e36d4049fb0fea195a8267c8dcd0824-Abstract-Conference.html}.

\bibitem[Lowe(2004)]{lowe2004sift}
Lowe, D.~G.
\newblock {Distinctive Image Features from Scale-Invariant Keypoints}.
\newblock \emph{International Journal of Computer Vision}, 60\penalty0
  (2):\penalty0 91--110, 2004.
\newblock URL \url{https://doi.org/10.1023/B:VISI.0000029664.99615.94}.

\bibitem[Mailthody et~al.(2019)Mailthody, Qureshi, Liang, Feng, {Garcia de
  Gonzalo}, Li, Franke, Xiong, Huang, and Hwu]{lee2024deepstore}
Mailthody, V.~S., Qureshi, Z., Liang, W., Feng, Z., {Garcia de Gonzalo}, S.,
  Li, Y., Franke, H., Xiong, J., Huang, J., and Hwu, W.
\newblock {DeepStore}: In-storage acceleration for intelligent queries.
\newblock In \emph{Proceedings of the 52nd Annual IEEE/ACM International
  Symposium on Microarchitecture (MICRO)}, pp.\  224--238, 2019.
\newblock URL \url{https://doi.org/10.1145/3352460.3358320}.

\bibitem[Malkov \& Yashunin(2020)Malkov and Yashunin]{malkov2020hnsw}
Malkov, Y.~A. and Yashunin, D.~A.
\newblock {Efficient and Robust Approximate Nearest Neighbor Search Using
  Hierarchical Navigable Small World Graphs}.
\newblock \emph{IEEE Transactions on Pattern Analysis and Machine
  Intelligence}, 42\penalty0 (4):\penalty0 824--836, 2020.
\newblock URL \url{https://doi.org/10.1109/TPAMI.2018.2889473}.

\bibitem[Muja \& Lowe(2014)Muja and Lowe]{muja2014flann}
Muja, M. and Lowe, D.~G.
\newblock {Scalable Nearest Neighbor Algorithms for High Dimensional Data}.
\newblock \emph{IEEE Transactions on Pattern Analysis and Machine
  Intelligence}, 36\penalty0 (11):\penalty0 2227--2240, 2014.
\newblock URL \url{https://doi.org/10.1109/tpami.2014.2321376}.

\bibitem[Nguyen et~al.(2016)Nguyen, Rosenberg, Song, Gao, Tiwary, Majumder, and
  Deng]{nguyen2016msmarco}
Nguyen, T., Rosenberg, M., Song, X., Gao, J., Tiwary, S., Majumder, R., and
  Deng, L.
\newblock {MS MARCO: A Human Generated MAchine Reading COmprehension Dataset}.
\newblock In \emph{Proceedings of the Workshop on Cognitive Computation:
  Integrating Neural and Symbolic Approaches, co-located with NeurIPS 2016},
  volume 1773 of \emph{CEUR Workshop Proceedings}, 2016.
\newblock URL \url{https://ceur-ws.org/Vol-1773/CoCoNIPS_2016_paper9.pdf}.

\bibitem[{NVIDIA}(2022)]{nvidia2022h100}
{NVIDIA}.
\newblock {NVIDIA} {H100} tensor core {GPU} architecture.
\newblock Whitepaper, 2022.
\newblock URL
  \url{https://resources.nvidia.com/en-us-hopper-architecture/nvidia-h100-tensor-c}.

\bibitem[{NVIDIA}(2025)]{nvidia2025cutlass}
{NVIDIA}.
\newblock {CUTLASS}: {CUDA} templates for linear algebra subroutines.
\newblock GitHub repository, 2025.
\newblock URL \url{https://github.com/NVIDIA/cutlass}.

\bibitem[{NVIDIA}(2026)]{nvidia2025cuvs}
{NVIDIA}.
\newblock {cuVS}: Vector search and clustering on the {GPU}.
\newblock GitHub repository, 2026.
\newblock URL \url{https://github.com/rapidsai/cuvs}.
\newblock Version 26.10.

\bibitem[Ootomo et~al.(2024)Ootomo, Naruse, Nolet, Wang, Feher, and
  Wang]{ootomo2024cagra}
Ootomo, H., Naruse, A., Nolet, C., Wang, R., Feher, T., and Wang, Y.
\newblock {CAGRA: Highly Parallel Graph Construction and Approximate Nearest
  Neighbor Search for GPUs}.
\newblock In \emph{Proceedings of the 40th IEEE International Conference on
  Data Engineering (ICDE)}, pp.\  4236--4247, 2024.
\newblock URL \url{https://doi.org/10.1109/ICDE60146.2024.00323}.

\bibitem[{OpenAI}(2025)]{openai2025gptoss}
{OpenAI}.
\newblock {gpt-oss-120b} \& {gpt-oss-20b} model card.
\newblock \emph{arXiv preprint arXiv:2508.10925}, 2025.
\newblock URL \url{https://arxiv.org/abs/2508.10925}.

\bibitem[Oquab et~al.(2024)Oquab, Darcet, Moutakanni, Vo, Szafraniec, Khalidov,
  Fernandez, Haziza, Massa, El-Nouby, Assran, Ballas, Galuba, Howes, Huang, Li,
  Misra, Rabbat, Sharma, Synnaeve, Xu, Jegou, Mairal, Labatut, Joulin, and
  Bojanowski]{oquab2024dinov2}
Oquab, M., Darcet, T., Moutakanni, T., Vo, H., Szafraniec, M., Khalidov, V.,
  Fernandez, P., Haziza, D., Massa, F., El-Nouby, A., Assran, M., Ballas, N.,
  Galuba, W., Howes, R., Huang, P.-Y., Li, S.-W., Misra, I., Rabbat, M.,
  Sharma, V., Synnaeve, G., Xu, H., Jegou, H., Mairal, J., Labatut, P., Joulin,
  A., and Bojanowski, P.
\newblock {DINOv2: Learning Robust Visual Features without Supervision}.
\newblock \emph{Transactions on Machine Learning Research}, 2024.
\newblock URL \url{https://openreview.net/forum?id=a68SUt6zFt}.

\bibitem[Paszke et~al.(2019)Paszke, Gross, Massa, Lerer, Bradbury, Chanan,
  Killeen, Lin, Gimelshein, Antiga, Desmaison, Köpf, Yang, DeVito, Raison,
  Tejani, Chilamkurthy, Steiner, Fang, Bai, and Chintala]{paszke2019pytorch}
Paszke, A., Gross, S., Massa, F., Lerer, A., Bradbury, J., Chanan, G., Killeen,
  T., Lin, Z., Gimelshein, N., Antiga, L., Desmaison, A., Köpf, A., Yang, E.,
  DeVito, Z., Raison, M., Tejani, A., Chilamkurthy, S., Steiner, B., Fang, L.,
  Bai, J., and Chintala, S.
\newblock {PyTorch: An Imperative Style, High-Performance Deep Learning
  Library}.
\newblock In \emph{Advances in Neural Information Processing Systems
  (NeurIPS)}, volume~32, 2019.
\newblock URL
  \url{https://proceedings.neurips.cc/paper/2019/hash/bdbca288fee7f92f2bfa9f7012727740-Abstract.html}.

\bibitem[Patel et~al.(2024)Patel, Kraft, Guestrin, and Zaharia]{patel2024acorn}
Patel, L., Kraft, P., Guestrin, C., and Zaharia, M.
\newblock {ACORN}: Performant and predicate-agnostic search over vector
  embeddings and structured data.
\newblock \emph{Proceedings of the ACM on Management of Data}, 2\penalty0
  (3):\penalty0 1--27, 2024.
\newblock URL \url{https://doi.org/10.1145/3654923}.

\bibitem[Peng et~al.(2026)Peng, Yang, Xie, Sun, Shou, Chen, and
  Chen]{peng2026svfusion}
Peng, Y., Yang, D., Xie, Z., Sun, J., Shou, L., Chen, K., and Chen, G.
\newblock {SVFusion}: A {CPU-GPU} co-processing architecture for large-scale
  real-time vector search.
\newblock \emph{Proceedings of the VLDB Endowment}, 19\penalty0 (5):\penalty0
  1074--1087, 2026.
\newblock URL \url{https://www.vldb.org/pvldb/vol19/p1074-yang.pdf}.

\bibitem[Qureshi et~al.(2023)Qureshi, Mailthody, Gelado, Min, Masood, Park,
  Xiong, Newburn, Vainbrand, Chung, Garland, Dally, and Hwu]{gao2023bam}
Qureshi, Z., Mailthody, V.~S., Gelado, I., Min, S., Masood, A., Park, J.,
  Xiong, J., Newburn, C.~J., Vainbrand, D., Chung, I.-H., Garland, M., Dally,
  W., and Hwu, W.
\newblock {GPU}-initiated on-demand high-throughput storage access in the {BaM}
  system architecture.
\newblock In \emph{Proceedings of the 28th ACM International Conference on
  Architectural Support for Programming Languages and Operating Systems
  (ASPLOS), Volume 2}, pp.\  325--339, 2023.
\newblock URL \url{https://doi.org/10.1145/3575693.3575748}.

\bibitem[{Qwen Team}(2026)]{qwen2026qwen38}
{Qwen Team}.
\newblock {Qwen3.8-Flash-Next}.
\newblock Hugging Face model repository, 2026.
\newblock URL \url{https://huggingface.co/Qwen/Qwen3.8-Flash-Next}.

\bibitem[Radford et~al.(2021)Radford, Kim, Hallacy, Ramesh, Goh, Agarwal,
  Sastry, Askell, Mishkin, Clark, Krueger, and Sutskever]{radford2021clip}
Radford, A., Kim, J.~W., Hallacy, C., Ramesh, A., Goh, G., Agarwal, S., Sastry,
  G., Askell, A., Mishkin, P., Clark, J., Krueger, G., and Sutskever, I.
\newblock {Learning Transferable Visual Models From Natural Language
  Supervision}.
\newblock In \emph{International Conference on Machine Learning (ICML)}, volume
  139 of \emph{Proceedings of Machine Learning Research}, pp.\  8748--8763,
  2021.
\newblock URL \url{https://proceedings.mlr.press/v139/radford21a.html}.

\bibitem[Rae et~al.(2020)Rae, Potapenko, Jayakumar, Hillier, and
  Lillicrap]{rae2020pg19}
Rae, J.~W., Potapenko, A., Jayakumar, S.~M., Hillier, C., and Lillicrap, T.~P.
\newblock Compressive transformers for long-range sequence modelling.
\newblock In \emph{International Conference on Learning Representations
  (ICLR)}, 2020.
\newblock URL \url{https://openreview.net/forum?id=SylKikSYDH}.

\bibitem[Reimers \& Gurevych(2019)Reimers and Gurevych]{reimers2019sbert}
Reimers, N. and Gurevych, I.
\newblock {Sentence-BERT: Sentence Embeddings using Siamese BERT-Networks}.
\newblock In \emph{Proceedings of the 2019 Conference on Empirical Methods in
  Natural Language Processing and the 9th International Joint Conference on
  Natural Language Processing (EMNLP-IJCNLP)}, pp.\  3982--3992, 2019.
\newblock \doi{10.18653/v1/D19-1410}.
\newblock URL \url{https://aclanthology.org/D19-1410/}.

\bibitem[Ren et~al.(2020)Ren, Zhang, and Li]{ren2020hmann}
Ren, J., Zhang, M., and Li, D.
\newblock {HM-ANN}: Efficient billion-point nearest neighbor search on
  heterogeneous memory.
\newblock In \emph{Advances in Neural Information Processing Systems
  (NeurIPS)}, volume~33, pp.\  10672--10684, 2020.
\newblock URL
  \url{https://proceedings.neurips.cc/paper/2020/hash/788d986905533aba051261497ecffcbb-Abstract.html}.

\bibitem[Shewchuk(1997)]{shewchuk1997adaptive}
Shewchuk, J.~R.
\newblock Adaptive precision floating-point arithmetic and fast robust
  geometric predicates.
\newblock \emph{Discrete \& Computational Geometry}, 18\penalty0 (3):\penalty0
  305--363, 1997.
\newblock URL \url{https://doi.org/10.1007/PL00009321}.

\bibitem[Shi et~al.(2026)Shi, Gao, Xia, Feh{\'e}r, and Long]{shi2026ivfrabitq}
Shi, J., Gao, J., Xia, J., Feh{\'e}r, T.~B., and Long, C.
\newblock {GPU}-native approximate nearest neighbor search with {IVF-RaBitQ}:
  Fast index build and search.
\newblock \emph{Proceedings of the VLDB Endowment}, 19\penalty0 (11):\penalty0
  3454--3467, 2026.
\newblock URL \url{https://doi.org/10.14778/3836663.3836701}.

\bibitem[Simhadri et~al.(2022)Simhadri, Williams, Aum{\"u}ller, Douze, Babenko,
  Baranchuk, Chen, Hosseini, Krishnaswamy, Srinivasa, Subramanya, and
  Wang]{simhadri2022bigann}
Simhadri, H.~V., Williams, G., Aum{\"u}ller, M., Douze, M., Babenko, A.,
  Baranchuk, D., Chen, Q., Hosseini, L., Krishnaswamy, R., Srinivasa, G.,
  Subramanya, S.~J., and Wang, J.
\newblock Results of the {NeurIPS'21} challenge on billion-scale approximate
  nearest neighbor search.
\newblock In \emph{Proceedings of the NeurIPS 2021 Competitions and
  Demonstrations Track}, volume 176 of \emph{Proceedings of Machine Learning
  Research}, pp.\  177--189, 2022.
\newblock URL \url{https://proceedings.mlr.press/v176/simhadri22a.html}.

\bibitem[Singh et~al.(2021)Singh, Subramanya, Krishnaswamy, and
  Simhadri]{singh2021freshdiskann}
Singh, A., Subramanya, S.~J., Krishnaswamy, R., and Simhadri, H.~V.
\newblock {FreshDiskANN}: A fast and accurate graph-based {ANN} index for
  streaming similarity search.
\newblock \emph{arXiv preprint arXiv:2105.09613}, 2021.
\newblock URL \url{https://arxiv.org/abs/2105.09613}.

\bibitem[Su et~al.(2024)Su, Ahmed, Lu, Pan, Bo, and Liu]{su2024roformer}
Su, J., Ahmed, M., Lu, Y., Pan, S., Bo, W., and Liu, Y.
\newblock {RoFormer}: Enhanced transformer with rotary position embedding.
\newblock \emph{Neurocomputing}, 568:\penalty0 127063, 2024.
\newblock \doi{10.1016/j.neucom.2023.127063}.
\newblock URL \url{https://doi.org/10.1016/j.neucom.2023.127063}.

\bibitem[Subramanya et~al.(2019)Subramanya, Devvrit, Simhadri, Krishnaswamy,
  and Kadekodi]{subramanya2019diskann}
Subramanya, S.~J., Devvrit, Simhadri, H.~V., Krishnaswamy, R., and Kadekodi, R.
\newblock {DiskANN: Fast Accurate Billion-point Nearest Neighbor Search on a
  Single Node}.
\newblock In \emph{Advances in Neural Information Processing Systems
  (NeurIPS)}, volume~32, 2019.
\newblock URL
  \url{https://papers.nips.cc/paper_files/paper/2019/hash/09853c7fb1d3f8ee67a61b6bf4a7f8e6-Abstract.html}.

\bibitem[Tang et~al.(2024)Tang, Zhao, Zhu, Xiao, Kasikci, and
  Han]{tang2024quest}
Tang, J., Zhao, Y., Zhu, K., Xiao, G., Kasikci, B., and Han, S.
\newblock Quest: Query-aware sparsity for efficient long-context {LLM}
  inference.
\newblock In \emph{International Conference on Machine Learning (ICML)}, volume
  235, pp.\  47901--47911, 2024.
\newblock URL \url{https://proceedings.mlr.press/v235/tang24l.html}.

\bibitem[{Tencent Hunyuan Team}(2026)]{tencent2026hy3}
{Tencent Hunyuan Team}.
\newblock {Hy3}.
\newblock Hugging Face model repository, 2026.
\newblock URL \url{https://huggingface.co/tencent/Hy3}.

\bibitem[Tian et~al.(2025)Tian, Liu, Tang, Xiao, Duan, Liao, Jin, Zhang, Zhu,
  and Zhang]{tian2025fusionanns}
Tian, B., Liu, H., Tang, Y., Xiao, S., Duan, Z., Liao, X., Jin, H., Zhang, X.,
  Zhu, J., and Zhang, Y.
\newblock Towards high-throughput and low-latency billion-scale vector search
  via {CPU/GPU} collaborative filtering and re-ranking.
\newblock In \emph{23rd USENIX Conference on File and Storage Technologies
  (FAST 25)}, pp.\  171--185, 2025.
\newblock URL
  \url{https://www.usenix.org/conference/fast25/presentation/tian-bing}.

\bibitem[Tillet et~al.(2019)Tillet, Kung, and Cox]{tillet2019triton}
Tillet, P., Kung, H.~T., and Cox, D.
\newblock {Triton: an intermediate language and compiler for tiled neural
  network computations}.
\newblock In \emph{Proceedings of the 3rd ACM SIGPLAN International Workshop on
  Machine Learning and Programming Languages}, pp.\  10--19, 2019.
\newblock URL \url{https://doi.org/10.1145/3315508.3329973}.

\bibitem[Venkatasubba et~al.(2025)Venkatasubba, Khan, Singh, Simhadri, and
  Vedurada]{venkatasubba2025bang}
Venkatasubba, K., Khan, S., Singh, S., Simhadri, H.~V., and Vedurada, J.
\newblock {BANG}: Billion-scale approximate nearest neighbour search using a
  single {GPU}.
\newblock \emph{IEEE Transactions on Big Data}, 11\penalty0 (6):\penalty0
  3142--3157, 2025.
\newblock URL \url{https://doi.org/10.1109/TBDATA.2025.3581085}.

\bibitem[Vidal~Ruiz(1986)]{vidal1986aesa}
Vidal~Ruiz, E.
\newblock {An algorithm for finding nearest neighbours in (approximately)
  constant average time}.
\newblock \emph{Pattern Recognition Letters}, 4\penalty0 (3):\penalty0
  145--157, 1986.
\newblock URL \url{https://doi.org/10.1016/0167-8655(86)90013-9}.

\bibitem[Vovk et~al.(2005)Vovk, Gammerman, and Shafer]{vovk2005conformal}
Vovk, V., Gammerman, A., and Shafer, G.
\newblock \emph{Algorithmic Learning in a Random World}.
\newblock Springer, Boston, MA, 2005.
\newblock \doi{10.1007/b106715}.
\newblock URL \url{https://doi.org/10.1007/b106715}.

\bibitem[Wang et~al.(2021)Wang, Yi, Guo, Jin, Xu, Li, Wang, Guo, Li, Xu, Yu,
  Yuan, Zou, Long, Cai, Li, Zhang, Mo, Gu, Jiang, Wei, and Xie]{wang2021milvus}
Wang, J., Yi, X., Guo, R., Jin, H., Xu, P., Li, S., Wang, X., Guo, X., Li, C.,
  Xu, X., Yu, K., Yuan, Y., Zou, Y., Long, J., Cai, Y., Li, Z., Zhang, Z., Mo,
  Y., Gu, J., Jiang, R., Wei, Y., and Xie, C.
\newblock Milvus: A purpose-built vector data management system.
\newblock In \emph{Proceedings of the 2021 ACM SIGMOD International Conference
  on Management of Data}, pp.\  2614--2627, 2021.
\newblock URL \url{https://doi.org/10.1145/3448016.3457550}.

\bibitem[Wang et~al.(2024)Wang, Xu, Yi, Wu, Peng, Ke, Gao, Xu, Guo, and
  Xie]{wang2024starling}
Wang, M., Xu, W., Yi, X., Wu, S., Peng, Z., Ke, X., Gao, Y., Xu, X., Guo, R.,
  and Xie, C.
\newblock Starling: An {I/O}-efficient disk-resident graph index framework for
  high-dimensional vector similarity search on data segment.
\newblock \emph{Proceedings of the ACM on Management of Data}, 2\penalty0
  (1):\penalty0 1--27, 2024.
\newblock URL \url{https://doi.org/10.1145/3639269}.

\bibitem[Weber et~al.(1998)Weber, Schek, and Blott]{weber1998vafile}
Weber, R., Schek, H.-J., and Blott, S.
\newblock A quantitative analysis and performance study for similarity-search
  methods in high-dimensional spaces.
\newblock In \emph{Proceedings of the 24th International Conference on Very
  Large Data Bases (VLDB)}, pp.\  194--205, 1998.
\newblock URL \url{http://www.vldb.org/conf/1998/p194.pdf}.

\bibitem[Wei et~al.(2020)Wei, Wu, Wang, Lou, Zhan, Li, and
  Cai]{wei2020analyticdbv}
Wei, C., Wu, B., Wang, S., Lou, R., Zhan, C., Li, F., and Cai, Y.
\newblock {AnalyticDB-V}: A hybrid analytical engine towards query fusion for
  structured and unstructured data.
\newblock \emph{Proceedings of the VLDB Endowment}, 13\penalty0 (12):\penalty0
  3152--3165, 2020.
\newblock \doi{10.14778/3415478.3415541}.
\newblock URL \url{https://doi.org/10.14778/3415478.3415541}.

\bibitem[Wu et~al.(2022)Wu, Rabe, Hutchins, and Szegedy]{wu2022memorizing}
Wu, Y., Rabe, M.~N., Hutchins, D., and Szegedy, C.
\newblock Memorizing transformers.
\newblock In \emph{International Conference on Learning Representations
  (ICLR)}, 2022.
\newblock URL \url{https://openreview.net/forum?id=TrjbxzRcnf-}.

\bibitem[Xiao et~al.(2024{\natexlab{a}})Xiao, Zhang, Han, Xiao, Lin, Zhang,
  Liu, and Sun]{xiao2024infllm}
Xiao, C., Zhang, P., Han, X., Xiao, G., Lin, Y., Zhang, Z., Liu, Z., and Sun,
  M.
\newblock {InfLLM}: Training-free long-context extrapolation for {LLMs} with an
  efficient context memory.
\newblock In \emph{Advances in Neural Information Processing Systems
  (NeurIPS)}, 2024{\natexlab{a}}.
\newblock URL \url{https://neurips.cc/virtual/2024/poster/94480}.

\bibitem[Xiao et~al.(2024{\natexlab{b}})Xiao, Tian, Chen, Han, and
  Lewis]{xiao2024streamingllm}
Xiao, G., Tian, Y., Chen, B., Han, S., and Lewis, M.
\newblock Efficient streaming language models with attention sinks.
\newblock In \emph{International Conference on Learning Representations
  (ICLR)}, 2024{\natexlab{b}}.
\newblock URL \url{https://arxiv.org/abs/2309.17453}.

\bibitem[Yang et~al.(2024)Yang, Yang, Zhang, Hui, Zheng, Yu, Li, Liu, Huang,
  Wei, Lin, Yang, Tu, Zhang, Yang, Yang, Zhou, Lin, Dang, Lu, Bao, Yang, Yu,
  Li, Xue, Zhang, Zhu, Men, Lin, Li, Tang, Xia, Ren, Ren, Fan, Su, Zhang, Wan,
  Liu, Cui, Zhang, and Qiu]{yang2024qwen25}
Yang, A., Yang, B., Zhang, B., Hui, B., Zheng, B., Yu, B., Li, C., Liu, D.,
  Huang, F., Wei, H., Lin, H., Yang, J., Tu, J., Zhang, J., Yang, J., Yang, J.,
  Zhou, J., Lin, J., Dang, K., Lu, K., Bao, K., Yang, K., Yu, L., Li, M., Xue,
  M., Zhang, P., Zhu, Q., Men, R., Lin, R., Li, T., Tang, T., Xia, T., Ren, X.,
  Ren, X., Fan, Y., Su, Y., Zhang, Y., Wan, Y., Liu, Y., Cui, Z., Zhang, Z.,
  and Qiu, Z.
\newblock {Qwen2.5} technical report.
\newblock \emph{arXiv preprint arXiv:2412.15115}, 2024.
\newblock URL \url{https://arxiv.org/abs/2412.15115}.

\bibitem[Yang et~al.(2025{\natexlab{a}})Yang, Li, Yang, Zhang, Hui, Zheng, Yu,
  Gao, Huang, Lv, Zheng, Liu, Zhou, Huang, Hu, Ge, Wei, Lin, Tang, Yang, Tu,
  Zhang, Yang, Yang, Zhou, Zhou, Lin, Dang, Bao, Yang, Yu, Deng, Li, Xue, Li,
  Zhang, Wang, Zhu, Men, Gao, Liu, Luo, Li, Tang, Yin, Ren, Wang, Zhang, Ren,
  Fan, Su, Zhang, Zhang, Wan, Liu, Wang, Cui, Zhang, Zhou, and
  Qiu]{yang2025qwen3}
Yang, A., Li, A., Yang, B., Zhang, B., Hui, B., Zheng, B., Yu, B., Gao, C.,
  Huang, C., Lv, C., Zheng, C., Liu, D., Zhou, F., Huang, F., Hu, F., Ge, H.,
  Wei, H., Lin, H., Tang, J., Yang, J., Tu, J., Zhang, J., Yang, J., Yang, J.,
  Zhou, J., Zhou, J., Lin, J., Dang, K., Bao, K., Yang, K., Yu, L., Deng, L.,
  Li, M., Xue, M., Li, M., Zhang, P., Wang, P., Zhu, Q., Men, R., Gao, R., Liu,
  S., Luo, S., Li, T., Tang, T., Yin, W., Ren, X., Wang, X., Zhang, X., Ren,
  X., Fan, Y., Su, Y., Zhang, Y., Zhang, Y., Wan, Y., Liu, Y., Wang, Z., Cui,
  Z., Zhang, Z., Zhou, Z., and Qiu, Z.
\newblock {Qwen3} technical report.
\newblock \emph{arXiv preprint arXiv:2505.09388}, 2025{\natexlab{a}}.
\newblock URL \url{https://arxiv.org/abs/2505.09388}.

\bibitem[Yang et~al.(2025{\natexlab{b}})Yang, Yu, Li, Liu, Huang, Huang, Jiang,
  Tu, Zhang, Zhou, Lin, Dang, Yang, Yu, Li, Sun, Zhu, Men, He, Xu, Yin, Yu,
  Qiu, Ren, Yang, Li, Xu, and Zhang]{yang2025qwen251m}
Yang, A., Yu, B., Li, C., Liu, D., Huang, F., Huang, H., Jiang, J., Tu, J.,
  Zhang, J., Zhou, J., Lin, J., Dang, K., Yang, K., Yu, L., Li, M., Sun, M.,
  Zhu, Q., Men, R., He, T., Xu, W., Yin, W., Yu, W., Qiu, X., Ren, X., Yang,
  X., Li, Y., Xu, Z., and Zhang, Z.
\newblock Qwen2.5-1m technical report.
\newblock \emph{arXiv preprint arXiv:2501.15383}, 2025{\natexlab{b}}.
\newblock URL \url{https://arxiv.org/abs/2501.15383}.

\bibitem[Ye et~al.(2025)Ye, Chen, Lai, Lin, Zhang, Wang, Chen, Kasikci, Grover,
  Krishnamurthy, and Ceze]{ye2025flashinfer}
Ye, Z., Chen, L., Lai, R., Lin, W., Zhang, Y., Wang, S., Chen, T., Kasikci, B.,
  Grover, V., Krishnamurthy, A., and Ceze, L.
\newblock {FlashInfer}: Efficient and customizable attention engine for {LLM}
  inference serving.
\newblock In \emph{Proceedings of Machine Learning and Systems (MLSys)},
  volume~7, 2025.
\newblock URL
  \url{https://proceedings.mlsys.org/paper_files/paper/2025/hash/dbf02b21d77409a2db30e56866a8ab3a-Abstract-Conference.html}.

\bibitem[Yianilos(1993)]{yianilos1993vptree}
Yianilos, P.~N.
\newblock Data structures and algorithms for nearest neighbor search in general
  metric spaces.
\newblock In \emph{Proceedings of the Fourth Annual ACM-SIAM Symposium on
  Discrete Algorithms (SODA)}, pp.\  311--321, 1993.
\newblock URL \url{https://dl.acm.org/doi/10.5555/313559.313789}.

\bibitem[Yu et~al.(2022)Yu, Jeong, Kim, Kim, and Chun]{yu2022orca}
Yu, G.-I., Jeong, J.~S., Kim, G.-W., Kim, S., and Chun, B.-G.
\newblock Orca: A distributed serving system for transformer-based generative
  models.
\newblock In \emph{16th USENIX Symposium on Operating Systems Design and
  Implementation (OSDI 22)}, pp.\  521--538, 2022.
\newblock URL \url{https://www.usenix.org/conference/osdi22/presentation/yu}.

\bibitem[Zhang et~al.(2023)Zhang, Xu, Chen, Sui, Xie, Cai, Chen, He, Yang,
  Yang, Yang, and Zhou]{zhang2023vbase}
Zhang, Q., Xu, S., Chen, Q., Sui, G., Xie, J., Cai, Z., Chen, Y., He, Y., Yang,
  Y., Yang, F., Yang, M., and Zhou, L.
\newblock {VBASE}: Unifying online vector similarity search and relational
  queries via relaxed monotonicity.
\newblock In \emph{17th USENIX Symposium on Operating Systems Design and
  Implementation (OSDI 23)}, pp.\  377--395, 2023.
\newblock URL
  \url{https://www.usenix.org/conference/osdi23/presentation/zhang-qianxi}.

\bibitem[Zhang et~al.(2024)Zhang, Liu, Huang, Liu, and Jin]{zhang2024rummy}
Zhang, Z., Liu, F., Huang, G., Liu, X., and Jin, X.
\newblock Fast vector query processing for large datasets beyond {GPU} memory
  with reordered pipelining.
\newblock In \emph{21st USENIX Symposium on Networked Systems Design and
  Implementation (NSDI 24)}, pp.\  23--40, 2024.
\newblock URL
  \url{https://www.usenix.org/conference/nsdi24/presentation/zhang-zili-pipelining}.

\bibitem[Zhao(2026)]{zhao2026certex}
Zhao, D.
\newblock {CertEx: Exact Visual Retrieval at Scale by Certified Exclusion}.
\newblock \emph{arXiv preprint arXiv:2610.21325}, 2026.
\newblock URL \url{https://arxiv.org/abs/2610.21325}.

\bibitem[Zhao et~al.(2020)Zhao, Tan, and Li]{zhao2020song}
Zhao, W., Tan, S., and Li, P.
\newblock {SONG: Approximate Nearest Neighbor Search on GPU}.
\newblock In \emph{2020 IEEE 36th International Conference on Data Engineering
  (ICDE)}, pp.\  1033--1044, 2020.
\newblock URL \url{https://doi.org/10.1109/icde48307.2020.00094}.

\end{thebibliography}
\bibliographystyle{mlsys2025}

\appendix

\section{Certified Screening Rows}
\label{supp:screen}

The propositions of this appendix build on one guarantee of screening, which Section~\ref{sec:bg:certex} describes and the screening stage of Section~\ref{sec:design:overview} relies on. This section restates it in the notation of \sysname. Consider a dataset of vectors $\mathbf{x} \in \mathbb{R}^D$, a center $\mathbf{c} \in \mathbb{R}^D$, and a matrix $\mathbf{V} \in \mathbb{R}^{m \times D}$ with orthonormal rows, which \sysname\ computes as the leading eigenvectors of the covariance of a sample of the dataset. CertEx~\cite{zhao2026certex} defines the lift
\begin{equation}
\lift(\mathbf{x}) = \bigl[\mathbf{V}(\mathbf{x}-\mathbf{c});\; \|(\mathbf{I}-\mathbf{V}^\top\mathbf{V})(\mathbf{x}-\mathbf{c})\|\bigr] \in \mathbb{R}^{m+1},
\label{eq:supp:lift}
\end{equation}
whose squared norm $a(\mathbf{x}) = \|\lift(\mathbf{x})\|^2$ equals $\|\mathbf{x}-\mathbf{c}\|^2$. Every query $\mathbf{q}$ and vector $\mathbf{x}$ thus satisfy $\|\mathbf{q}-\mathbf{x}\|^2 \ge a(\mathbf{q})+a(\mathbf{x})-2\langle\lift(\mathbf{q}),\lift(\mathbf{x})\rangle$, because the Cauchy-Schwarz inequality bounds the inner product of the two residuals by the product of their norms. Throughout this appendix, $U(\mathbf{q})$ and $r_k(\mathbf{q})$ denote squared distances, which leaves the statements of the paper unchanged, as squaring preserves the order of distances.

\sysname\ evaluates this bound in half precision with the rounding certificate of CertEx~\cite{zhao2026certex}. Every stored vector becomes a screening row of $m+1$ half-precision lifted coordinates, scaled by a common factor $1/\alpha$ so that every lifted row has norm at most 64, followed by four certificate coordinates. Two certificate coordinates hold rounded-up bounds on the norm of the row and on its rounding error, while the other two hold $-a(\mathbf{x})/(2\alpha^2)$ as a sum of two half-precision numbers. With the matching coordinates in the query row, one tensor-core product $v(\mathbf{q},\mathbf{x})$ with fp32 accumulation evaluates the scaled bound together with its error terms. A vector survives if and only if $v(\mathbf{q},\mathbf{x}) \ge \theta(\mathbf{q})$, where
\begin{equation}
\theta(\mathbf{q}) = \mathrm{down}\Bigl(\frac{a(\mathbf{q}) - U(\mathbf{q})}{2\alpha^2} - \mu(\mathbf{q})\Bigr)
\label{eq:supp:theta}
\end{equation}
and the margin $\mu(\mathbf{q})$ covers the remaining error terms through the maxima of the certificate coordinates over the dataset. For a survivor, the computed product yields the certified lower bound
\begin{equation}
\lambda(\mathbf{q},\mathbf{x}) = \bigl(a(\mathbf{q})/\alpha^2 - 2(v(\mathbf{q},\mathbf{x}) + \epsilon(\mathbf{q}))\bigr)\alpha^2,
\label{eq:supp:lambda}
\end{equation}
where $\epsilon(\mathbf{q})$ collects the accumulation error and the margin.

\begin{proposition}[Certified Screening]
\label{prop:supp:screen}
If the certificate maxima cover every stored row and every product accumulates its terms in one sequential chain in single precision, then every vector that fails the test $v(\mathbf{q},\mathbf{x})\ge\theta(\mathbf{q})$ satisfies $\|\mathbf{q}-\mathbf{x}\|^2 > U(\mathbf{q})$, whereas every survivor satisfies $\lambda(\mathbf{q},\mathbf{x}) \le \|\mathbf{q}-\mathbf{x}\|^2$.
\end{proposition}

Proposition~\ref{prop:supp:screen} restates the guarantee of CertEx~\cite{zhao2026certex}, whose error analysis follows the standard analysis of floating-point dot products~\cite{higham2002accuracy} and of tensor-core arithmetic~\cite{blanchard2020tensorcore}. Its accumulation condition constrains the kernels of Section~\ref{sec:impl:kernels}. Specifically, the error constant $\gamma = 16\,L\,u$ with $u = 2^{-24}$ bounds a sequential fp32 accumulation over a row of $L$ padded halves. Every product thus accumulates in one chain of tensor-core instructions, since \sysname\ never splits the reduction over the row length across thread blocks.

\section{Correctness of Thresholds, Updates, and Inner Products}
\label{supp:correct}

The paper states three guarantees of the design in prose, which this section proves. The first concerns the partition threshold of Section~\ref{sec:design:ivf}, the second the insertions and deletions of Section~\ref{sec:design:updates}, and the third the inner-product screening of Section~\ref{sec:design:mips}. Each proof uses only the lower bound of Section~\ref{supp:screen} and the certificate of Proposition~\ref{prop:supp:screen}.

\begin{proposition}[Partition Threshold]
For a query $\mathbf{q}$ and any $k$ distinct stored vectors $\mathbf{x}_1, \dots, \mathbf{x}_k$, the true $k$-th nearest distance satisfies $r_k(\mathbf{q}) \le \max_i \|\mathbf{q} - \mathbf{x}_i\|$. Consequently, screening under $\min(U_0, U_1)$ never excludes a true $k$ nearest neighbor of $\mathbf{q}$.
\label{prop:ivf_termination}
\end{proposition}

\begin{proof}
The $k$-th nearest distance $r_k(\mathbf{q})$ is the smallest radius of a ball around $\mathbf{q}$ that contains $k$ stored vectors. The ball of radius $\max_i\|\mathbf{q}-\mathbf{x}_i\|$ contains the $k$ distinct vectors $\mathbf{x}_1,\dots,\mathbf{x}_k$, hence $r_k(\mathbf{q})\le\max_i\|\mathbf{q}-\mathbf{x}_i\|$. The threshold $U_0$ is the $k$-th smallest exact distance among distinct live anchors, which equals this maximum for the $k$ nearest anchors. Similarly, $U_1$ equals this maximum for the $k$ nearest of the $\rho k$ distinct candidates. Both thresholds and therefore their minimum bound $r_k(\mathbf{q})$ from above. Every true neighbor $\mathbf{x}$ then satisfies $\|\lift(\mathbf{q})-\lift(\mathbf{x})\|\le\|\mathbf{q}-\mathbf{x}\|\le\min(U_0,U_1)$, since the lifted distance never exceeds the true distance. The rounding certificate of CertEx~\cite{zhao2026certex} keeps every vector whose lifted distance is at most the threshold, which includes the true neighbor $\mathbf{x}$.
\end{proof}

\begin{proposition}[Update Correctness Invariant]
For a fixed orthonormal projection and center, the lifted lower bound holds for every stored vector. If the maxima of the rounding certificate cover every stored row, the certified test never excludes a true neighbor among the live vectors. Hence, insertions with the existing projection and deletions that disable rows and anchors preserve exactness.
\label{prop:update_invariant}
\end{proposition}

\begin{proof}
The lifted lower bound $\|\lift(\mathbf{q})-\lift(\mathbf{x})\|\le\|\mathbf{q}-\mathbf{x}\|$ of CertEx~\cite{zhao2026certex} holds for every orthonormal projection and every center, as its proof uses only the orthogonal split of $\mathbf{q}-\mathbf{c}$ and $\mathbf{x}-\mathbf{c}$ into the span of $\mathbf{V}$ and its complement. An insertion lifts the new vectors with the existing projection and center, for which the bound also holds. The rounding certificate bounds the error of the half-precision product through the database maxima of the certificate coordinates. Because an insertion raises these maxima to cover its rows before any search reads them, the certified test keeps every live vector within the threshold, as in a static dataset. A deleted row meets the constant coordinate of every query with the most negative half-precision value. Its product thus falls below every threshold, which excludes the row from every screening. A deleted anchor no longer enters the threshold, which thus remains the $k$-th smallest exact distance among live vectors and bounds the $k$-th distance among the live vectors from above by Proposition~\ref{prop:ivf_termination}.
\end{proof}

For inner-product queries, Section~\ref{sec:design:mips} lifts every vector around the origin as $\ell(\mathbf{x}) = [\mathbf{V}\mathbf{x};\, \|(\mathbf{I} - \mathbf{V}^\top \mathbf{V})\mathbf{x}\|]$ and screens the score $s(\mathbf{q}, \mathbf{x}) = \|\mathbf{q}\|^2 - 2\langle \mathbf{q}, \mathbf{x} \rangle$ by the lower bound
\begin{equation}
s(\mathbf{q}, \mathbf{x}) \ge \|\ell(\mathbf{q})\|^2 - 2\langle \ell(\mathbf{q}), \ell(\mathbf{x}) \rangle .
\label{eq:mips_lb}
\end{equation}

\begin{proposition}[Exact Inner-Product Screening]
Consider any orthonormal $\mathbf{V}$ and any $k$ distinct vectors whose $k$-th smallest score is $T$. Every vector whose bound in Eq.~\eqref{eq:mips_lb} exceeds $T$ lies outside the $k$ vectors of largest inner product with $\mathbf{q}$ among all stored vectors.
\label{prop:mips}
\end{proposition}

\begin{proof}
Define the residual $\mathbf{r}_{\mathbf{x}}\coloneqq(\mathbf{I}-\mathbf{V}^\top\mathbf{V})\mathbf{x}$, which gives $\ell(\mathbf{x})=[\mathbf{V}\mathbf{x};\,\|\mathbf{r}_{\mathbf{x}}\|]$ and likewise $\ell(\mathbf{q})$. Because the rows of $\mathbf{V}$ are orthonormal, $\mathbf{x}=\mathbf{V}^\top\mathbf{V}\mathbf{x}+\mathbf{r}_{\mathbf{x}}$ splits $\mathbf{x}$ into orthogonal parts, which gives $\langle\mathbf{q},\mathbf{x}\rangle=\langle\mathbf{V}\mathbf{q},\mathbf{V}\mathbf{x}\rangle+\langle\mathbf{r}_{\mathbf{q}},\mathbf{r}_{\mathbf{x}}\rangle$ and $\|\mathbf{q}\|^2=\|\ell(\mathbf{q})\|^2$. The Cauchy-Schwarz inequality bounds $\langle\mathbf{r}_{\mathbf{q}},\mathbf{r}_{\mathbf{x}}\rangle$ by $\|\mathbf{r}_{\mathbf{q}}\|\|\mathbf{r}_{\mathbf{x}}\|$, hence $\langle\mathbf{q},\mathbf{x}\rangle\le\langle\ell(\mathbf{q}),\ell(\mathbf{x})\rangle$, which yields Eq.~\eqref{eq:mips_lb}. The $k$ given vectors show that the $k$-th smallest score over the dataset is at most $T$. A vector whose bound exceeds $T$ has a score above $T$ and hence lies outside the $k$ vectors of smallest score. These vectors are the $k$ vectors of largest inner product, since the score decreases as the inner product grows.
\end{proof}

The inner-product bound of Section~\ref{sec:design:mips} has no term that depends on the vector alone. The screening rows thus keep their layout and their rounding certificate, with the two coordinates of $-a(\mathbf{x})/2$ set to zero. The threshold $T$ comes from the exact scores of the anchors, as in the Euclidean case. Verification then ranks the survivors by $s(\mathbf{q},\mathbf{x})$ in double precision and returns the exact top-$k$ by inner product.

\section{Implementation Details}
\label{supp:impl}

\subsection{GPU Screening Kernels}
\label{sec:impl:kernels}

On Hopper GPUs~\cite{nvidia2022h100}, every thread block of the screening kernel of Section~\ref{sec:design:overview} keeps a tile of 128 query rows in shared memory and walks every $G$-th tile of database rows, where $G$ is the number of blocks that share the query tile. The grid holds at most two resident blocks per streaming multiprocessor, as one more block would start a second wave and nearly double the pass time. The Tensor Memory Accelerator (TMA) loads the row tiles through a three-stage pipeline, from which warpgroup matrix-multiply instructions accumulate their half-precision products in single precision with the layouts of CUTLASS~\cite{nvidia2025cutlass}. The reduction over the row runs in one sequential loop in the block, which gives every product the accumulation order that Proposition~\ref{prop:supp:screen} assumes.

Each thread of the epilogue collects the survivors of its accumulator fragment in a bit mask and reserves space for them in the output segment of its block with one shared-memory atomic. Blocks thus never contend for a global counter. For rows of 128 and 256 halves, \sysname\ runs a warp-specialized variant, in which a producer warpgroup streams row tiles with TMA and two or three consumer warpgroups take the tiles in turn. The epilogue of one consumer then overlaps the tensor-core work of the others. Rows of 64 halves give each consumer too short a product to hide a heavy epilogue, and the cooperative kernel therefore serves them up to $2^{25}$ rows.

The routed variant takes a list of work items, each of which pairs a tile of queries with a tile of rows from one contiguous partition that TMA loads at any row offset. The same routed kernel screens the label partitions of filtered search. On Ampere GPUs such as the A100, \sysname\ runs kernels with the same epilogue that issue mma.sync instructions and load tiles with asynchronous copies in place of TMA.

\subsection{Verification from Host Memory}
\label{sec:impl:memory}

For the verification of Section~\ref{sec:design:tiered_mem}, the GPU reads the raw vectors from pinned host memory through unified virtual addressing, which spares verification any staging of rows on the device. The sort keys of Section~\ref{supp:ordered} hold the query index in their high bits and the order-preserving bit pattern of the lower bound in their low bits, which a radix sort~\cite{adinets2022onesweep} orders in one pass. Every temporary buffer of a batch comes from a stream-ordered allocator, since an ordinary device free synchronizes the whole GPU and would stall the pipeline of a batch at every buffer.

In round $r$ of verification, every unfinished query takes its next $32 \cdot 2^r$ survivors whose lower bounds lie below its current $k$-th exact distance. One flat launch computes the exact distances of all these pairs in double precision, with eight row reads of each warp in flight. A second launch merges the distances of each query into its top-$k$. A query finishes when its window stops at a lower bound above its $k$-th distance or at its last survivor.

\subsection{Updates and Interface}
\label{sec:impl:api}

For the updates of Section~\ref{sec:design:updates}, a deletion sets the coordinate of the screening row that meets the constant coordinate of every query row to the most negative half-precision value, which pushes the product of the row below every threshold. The deletion also marks the vector, which then no longer sets a threshold as an anchor. Searches hold a reader-writer lock in shared mode for the duration of a batch, while an update takes it in exclusive mode only while it publishes the new row count and the certificate maxima.

The interface of \sysname\ (Table~\ref{tab:api}) follows the build, extend, remove, and search calls of cuVS~\cite{nvidia2025cuvs}. A search takes a batch of queries, the number $k$ of neighbors, and optionally one label per query or a bitset filter. The index ranks the vectors by Euclidean distance, by cosine similarity, or by inner product. For sparse attention, a key cache keeps one index per KV head and returns the exact top-$k$ either in the layout of \texttt{torch.topk}~\cite{paszke2019pytorch} or as the int32 positions of the top-$k$ transforms of FlashInfer~\cite{ye2025flashinfer}. The engine chooses the number of projected dimensions and the anchor stride with the cost model of Section~\ref{sec:design:tuning} at build time.

\section{Per-Partition Local Projections}
\label{supp:local}

The local bounds of Section~\ref{sec:design:local_bounds} replace the global lift inside every partition, which requires the certificate to hold per partition. Every partition $\mathcal{P}_j$ receives its own center $\mathbf{c}_j$, its own orthonormal directions $\mathbf{V}_j$, its own scale $\alpha_j$, and its own certificate maxima, with which its rows are encoded. A search lifts every query once per partition with a batched matrix product. The search then sets the threshold of each pair of query and partition from the common $U(\mathbf{q})$ with the scale and maxima of the partition, after which one routed pass screens every partition with its own query rows.

\begin{proposition}[Local Projections]
\label{prop:supp:local}
If the rows of every partition are encoded with the projection and certificate maxima of that partition, and the queries of the routed pass for the partition are lifted with the same projection, then Proposition~\ref{prop:supp:screen} holds for every query of the routed pass and every row of the partition.
\end{proposition}

\begin{proof}
The lifted bound of Eq.~\eqref{eq:supp:lift} holds for every orthonormal $\mathbf{V}$ and every center, and in particular for $(\mathbf{c}_j,\mathbf{V}_j)$. The rounding certificate of a pair bounds the error of its product with the maxima of the rows that the query meets, which are the maxima of partition $\mathcal{P}_j$. Both premises of Proposition~\ref{prop:supp:screen} hence hold within each partition, where every pair meets the same $U(\mathbf{q})$.
\end{proof}

The anchors keep rows of the global projection, because the anchor stage compares every query with every anchor. Vectors added after the build likewise keep the global projection. The routed pass covers the partitions, and the plain pass the remaining rows, which subjects every vector to one certified test. Under a filter, every admitted neighbor still survives, as the threshold comes from admitted anchors. Verification then drops the survivors that the filter rejects, after which the remaining top-$k$ is the exact filtered answer.

\section{Compact int8 Screening Rows}
\label{supp:int8}

Half-precision screening rows take $2(m+5)$ bytes per vector, while int8 codes with one scale per row take about half of that at the same width, which extends the dataset that one GPU holds in the tiered placement of Section~\ref{sec:design:tiered_mem}. Every lifted vector stores its projection coordinates $\mathbf{z}_x = s_x(\mathbf{c}_x + \mathbf{e}_x)$ as int8 codes $\mathbf{c}_x$ with the scale $s_x = \max_j |z_{x,j}|/127$, which keeps $|e_{x,j}| \le 1/2$. Each row also stores $a(\mathbf{x})/2$, $s_x$, $\|\mathbf{e}_x\|$, $\|\mathbf{c}_x\|$, and the residual norm in single precision. The tensor cores accumulate $\langle\mathbf{c}_q,\mathbf{c}_x\rangle$ exactly in int32, and the value converts exactly to single precision because $127^2 m < 2^{24}$.

The Cauchy-Schwarz inequality bounds the remaining terms of the product of the projection coordinates, which gives
\begin{equation}
\begin{split}
\langle \mathbf{z}_q, \mathbf{z}_x\rangle \le s_q s_x \bigl(&\langle\mathbf{c}_q,\mathbf{c}_x\rangle + \|\mathbf{e}_x\|(\|\mathbf{c}_q\| + \|\mathbf{e}_q\|)\\
&+ \|\mathbf{e}_q\|\|\mathbf{c}_x\|\bigr).
\end{split}
\label{eq:supp:int8}
\end{equation}
Consider the lower bound $\beta(\mathbf{q},\mathbf{x})$ that results from the lifted bound with Eq.~\eqref{eq:supp:int8} in place of $\langle \mathbf{z}_q,\mathbf{z}_x\rangle$, and its single-precision evaluation $\tilde\beta$.

\begin{proposition}[int8 Rows]
\label{prop:supp:int8}
If the pass keeps every pair with $\tilde\beta \le U(\mathbf{q}) + \eta(\mathbf{q})$, where $\eta(\mathbf{q}) = 64 u (\sqrt{a(\mathbf{q})} + \sqrt{\max_x a(\mathbf{x})})^2$, then every excluded vector has $\|\mathbf{q}-\mathbf{x}\|^2 > U(\mathbf{q})$, whereas $\tilde\beta - \eta(\mathbf{q})$ bounds the squared distance of every survivor from below.
\end{proposition}

\begin{proof}
Eq.~\eqref{eq:supp:int8} bounds $\langle \mathbf{z}_q,\mathbf{z}_x\rangle$ from above, hence $\beta(\mathbf{q},\mathbf{x}) \le \|\lift(\mathbf{q})-\lift(\mathbf{x})\|^2 \le \|\mathbf{q}-\mathbf{x}\|^2$. The single-precision evaluation of $\beta$ takes at most 16 rounded operations per term, whose terms sum in magnitude to at most $2.6(\|\lift(\mathbf{q})\|+\|\lift(\mathbf{x})\|)^2$. With these counts, the standard error analysis of floating-point sums~\cite{higham2002accuracy} gives $|\tilde\beta - \beta| \le \eta(\mathbf{q})$. An excluded vector therefore has $\beta > \tilde\beta - \eta(\mathbf{q}) > U(\mathbf{q})$, whereas every survivor has $\tilde\beta - \eta(\mathbf{q}) \le \beta \le \|\mathbf{q}-\mathbf{x}\|^2$.
\end{proof}

The int8 rows return the same exact neighbors as half-precision rows, with the additional survivors that their quantization admits. The anchors keep half-precision rows, which serve only to choose the anchors whose exact distances form $U(\mathbf{q})$. Since the int8 bound only admits more survivors, Proposition~\ref{prop:supp:ordered} applies unchanged and the search returns the exact top-$k$.

\section{Ordered Verification}
\label{supp:ordered}

Ordered verification (Section~\ref{sec:design:tiered_mem}) reads the raw vector of a survivor only while the survivor can still enter the top $k$, which bounds the reads from host memory by the tightness of the certified lower bounds. A kernel computes $\lambda(\mathbf{q},\mathbf{x})$ of Eq.~\eqref{eq:supp:lambda} for every survivor and rounds it down to single precision. The kernel then forms a key that concatenates the query index with the order-preserving bit pattern of the bound, with which one radix sort orders the survivors by query and then by bound. A binary search finds the first survivor of every query. The implementation runs this verification in rounds over all queries of a batch (Section~\ref{sec:impl:memory}). The $k$-th distance at the start of a round is never below the one that a sequential verification would use at the same survivor. As a result, the rounds verify every survivor that sequential order verifies and return the same top-$k$.

\begin{proposition}[Ordered Verification]
\label{prop:supp:ordered}
If $U(\mathbf{q}) \ge r_k(\mathbf{q})$ and Proposition~\ref{prop:supp:screen} holds, the verification that stops at the first survivor whose bound exceeds the current $k$-th exact distance returns $k$ vectors of smallest distance.
\end{proposition}

\begin{proof}
Every vector $\mathbf{x}$ with $\|\mathbf{q}-\mathbf{x}\|^2 \le r_k(\mathbf{q})$ satisfies $\|\mathbf{q}-\mathbf{x}\|^2 \le U(\mathbf{q})$ and therefore survives by Proposition~\ref{prop:supp:screen}. Consider the moment when the verification stops with current $k$-th exact distance $b$. Every unverified survivor has a bound of at least the bound that triggered the stop, which exceeds $b$. Proposition~\ref{prop:supp:screen} then gives it a distance above $b$. As a result, an unverified survivor cannot replace any of the current $k$ vectors, whose distances are at most $b$. The current $k$ vectors are thus $k$ vectors of smallest distance among all survivors, and hence among all vectors.
\end{proof}

Proposition~\ref{prop:supp:ordered} bounds the number of exact distances by the survivors whose bound lies below the $k$-th distance at the start of the last round. Consequently, a query whose answer settles within its first 32 survivors reads at most 32 raw rows, while a survivor whose bound lies far above the $k$-th distance costs one key in the sort and no read of its vector. This property explains why the tiered layout of Section~\ref{sec:design:tiered_mem} keeps its throughput: on BIGANN-500M~\cite{jegou2011bigann}, verification reads only 699 of the 5,593 survivors per query.

\section{Multi-GPU Shared Threshold}
\label{supp:shards}

A dataset that exceeds one GPU splits into contiguous shards with one index per device, over which a search runs in two rounds (Section~\ref{sec:design:sharding}). In the first round, every shard computes the threshold $U_s(\mathbf{q})$ of every query from its own anchors. In the second round, every shard searches under the smallest threshold over all shards and returns the exact $k$ nearest vectors of its own dataset within that threshold, which the host merges by exact distance. Since every shard skips its own threshold stage in the second round, the two rounds cost little beyond a single-GPU search. Only $k$ identifier and distance pairs per shard and query cross the interconnect, which keeps the merge independent of the size of the shards. \sysname\ builds the shard indexes in parallel, with one host thread driving each device, and on four GPUs the host merges the answers of 10,000 queries in 16.7~ms.

\begin{proposition}[Shared Threshold]
\label{prop:supp:shards}
The merge of the shard answers under the threshold $\min_s U_s(\mathbf{q})$ returns $k$ vectors of smallest distance over all shards.
\end{proposition}

\begin{proof}
Each $U_s(\mathbf{q})$ is the $k$-th exact distance of $k$ distinct vectors of the dataset, hence $\min_s U_s(\mathbf{q}) \ge r_k(\mathbf{q})$ for the $k$-th distance $r_k(\mathbf{q})$ over all shards by Proposition~\ref{prop:ivf_termination}. Every vector among the $k$ nearest over all shards therefore lies within the shared threshold and among the $k$ nearest vectors of its own shard. Each shard returns its exact $k$ nearest vectors within the threshold by Proposition~\ref{prop:supp:ordered}. The merge of these sets therefore contains the $k$ nearest vectors.
\end{proof}

\section{Evaluation of the Cost Model}
\label{sec:eval:tuning}

The cost model of Section~\ref{sec:design:tuning} predicts the time of a batch as
\begin{equation}
\begin{split}
T(m,s,B) ={}& \alpha_{\text{scr}}\,B\,N_{\text{p}}(m+5) + \alpha_{\text{anc}}\,B\,\frac{N_{\text{p}}}{s}\\
&+ \alpha_{\text{ver}}\,B\,S(m,s),
\end{split}
\label{eq:cost_model}
\end{equation}
where $N_{\text{p}}$ is the number of vectors in the probed partitions, $S(m,s)$ is the expected number of survivors per query, and $\alpha_{\text{ver}}$ depends on the memory tier of the full-precision vectors.

We evaluate the cost model of Section~\ref{sec:design:tuning} on BIGANN~\cite{jegou2011bigann} from 10M to 500M vectors with the tiered layout, where tuning from prefixes of 1M and 4M vectors takes 4.9 to 5.2~s per dataset size. Since every candidate configuration is exact, a wrong choice costs only throughput.

\begin{figure}[t]
\centering
\includegraphics[width=\columnwidth]{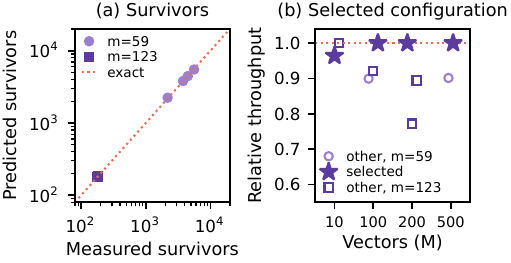}
\caption{Cost-model predictions and selections on BIGANN.}
\label{fig:eval_tuning}
\end{figure}

Figure~\ref{fig:eval_tuning} shows that the model predicts the survivors per query within 2.7\% over this 50$\times$ range of dataset sizes (Figure~\ref{fig:eval_tuning}(a)) and selects the fastest measured configuration from 100M to 500M vectors (Figure~\ref{fig:eval_tuning}(b)). The selected width shrinks from 123 at 10M and 100M to 59 at 200M and 500M, which gives 11.4k against 10.2k queries per second at 200M, although the rows of width 123 still fit in device memory. At 10M, the selected anchor stride of 16 lies within 4\% of the faster stride of 32. The predicted throughput of the selected configurations lies 1\% to 11\% above the measured one.

\section{Configuration Effects}
\label{supp:tuning}

Measurements of the CertEx engine~\cite{zhao2026certex} at every evaluated width on 21 dataset configurations reveal three effects that a static choice cannot capture, which motivate the tuner of Section~\ref{sec:design:tuning}. First, the throughput is unimodal in $m$ with a flat top. Widths below the best one lose up to a factor of 5 as the looser bound multiplies the survivors, while widths above it lose only 5\% to 35\%. A static rule that takes the smallest tile-aligned width capturing 95\% of the variance of the database reaches at least 0.87 of the best throughput and selects the best width on 12 of the 21 configurations. Second, the best width shrinks as the dataset grows, since the screening pass scales with $N(m+5)$ while the verification cost per query depends only on the number of survivors. On BIGANN~\cite{jegou2011bigann}, the width 123 is fastest up to 20M vectors, whereas the width 59 is fastest from 50M vectors on, which no rule based on the variance alone can predict. Third, the variance of the database does not describe queries from another distribution. For CLIP~\cite{radford2021clip} text queries against CLIP image embeddings, the width of the 95\% rule admits more than $2\times10^8$ survivors per batch of 10{,}000 queries and overflows the survivor buffer, whereas the next width is the fastest. A batch whose survivors exceed the staging buffer is re-executed for each half of its queries, which keeps every answer exact even when the survivor count was underestimated. The re-executed passes keep the shape of the full batch and give the inactive queries an infinite threshold, as a new batch shape would re-tune the kernel during the query and cost several seconds.

\section{Filtered Search and Label Partitions}
\label{supp:labels}

Filtered search admits for every query only the vectors that a predicate accepts. An exact filtered answer is thus the exact top-$k$ over these admitted vectors, which both ways of Section~\ref{sec:design:filtered} return. The first way applies the predicate only to the pairs that pass the threshold, after the anchor stage has applied the same predicate to the anchors. The threshold of every query thus comes from admitted vectors alone. Proposition~\ref{prop:ivf_termination} and Proposition~\ref{prop:supp:ordered} with the admitted vectors of each query then give the exact filtered answer, in which every entry beyond the admitted vectors holds the id $-1$ in place of a neighbor.

The second way sorts a copy of the screening rows by label and maps every row of the copy to the id of its vector. A search with query labels sorts each batch of queries by label and routes every tile of queries to the rows of its label. This routed kernel screens only tiles that pair queries of one label with rows of the same label and masks the rows of a loaded tile that lie outside the label. Vectors inserted after partitioning form a tail, which the first way screens. The survivors of the tail then join those of the routed pass.

\begin{proposition}[Label Partitions]
\label{prop:supp:labels}
A search with query labels over a label partition and its tail returns, for every query, $k$ live vectors of its label of smallest distance.
\end{proposition}

\begin{proof}
The routed pass screens every row of the label of each query in the partition, while the first way screens every row of the tail under the label predicate. Every admitted vector of a query thus meets the certified test exactly once. The masks also keep no pair of different labels. The threshold comes from live anchors of the label of the query and bounds the $k$-th admitted distance by Proposition~\ref{prop:ivf_termination}. Since the survivors map to vector ids before verification, Propositions~\ref{prop:update_invariant} and~\ref{prop:supp:ordered} applied to the live vectors of the label of each query then give the claim.
\end{proof}

\section{Exact Cosine Similarity}
\label{supp:cosine}

Cosine similarity (Section~\ref{sec:design:mips}) needs a separate argument, as \sysname\ returns the exact cosine of the vectors as given while the screening rows lift normalized copies whose rounding perturbs them. The score of a pair is $s(\mathbf{q},\mathbf{x}) = 2 - 2\langle\mathbf{q},\mathbf{x}\rangle/(\|\mathbf{q}\|\|\mathbf{x}\|)$, the squared distance of the exact unit vectors, which \sysname\ computes in double precision from the stored vectors. The screening rows instead lift the float32-normalized copies $\mathbf{q}'$ and $\mathbf{x}'$, whose rounding moves each of them from its exact unit vector by at most $\eta_1 = (D+16)u$.

The triangle inequality bounds the difference between $\|\mathbf{q}'-\mathbf{x}'\|$ and the exact unit distance by $\eta = 2\eta_1$. Because $\|\mathbf{q}'-\mathbf{x}'\| \le 2+\eta$, the exact score is at least $\|\mathbf{q}'-\mathbf{x}'\|^2 - \delta$ with $\delta = 2\eta(2+\eta)$. \sysname\ therefore adds $\delta$ to $U(\mathbf{q})$ in Eq.~\eqref{eq:supp:theta} and subtracts it from $\lambda(\mathbf{q},\mathbf{x})$ in Eq.~\eqref{eq:supp:lambda}.

\begin{proposition}[Exact Cosine]
\label{prop:supp:cosine}
With the slack $\delta$, every vector that fails the certified test has a cosine score above $U(\mathbf{q})$, whereas every survivor satisfies $\lambda(\mathbf{q},\mathbf{x}) - \delta \le s(\mathbf{q},\mathbf{x})$.
\end{proposition}

\begin{proof}
A vector that fails the test satisfies $\|\mathbf{q}'-\mathbf{x}'\|^2 > U(\mathbf{q})+\delta$ by Proposition~\ref{prop:supp:screen} applied to the normalized copies with the threshold $U(\mathbf{q})+\delta$. Its exact score is at least $\|\mathbf{q}'-\mathbf{x}'\|^2-\delta$ and thus exceeds $U(\mathbf{q})$. Every survivor satisfies $\lambda(\mathbf{q},\mathbf{x}) \le \|\mathbf{q}'-\mathbf{x}'\|^2$ by the same proposition, which together with the lower bound on the exact score gives $\lambda(\mathbf{q},\mathbf{x})-\delta \le s(\mathbf{q},\mathbf{x})$.
\end{proof}

Proposition~\ref{prop:supp:cosine} supplies the two properties that Propositions~\ref{prop:ivf_termination} and~\ref{prop:supp:ordered} use, namely exclusion above the threshold and a lower bound for every survivor. Since the thresholds of cosine search come from exact scores of anchors and candidates, both propositions apply with $s$ in place of the squared distance. The inner-product case needs no slack, because its lift of Section~\ref{sec:design:mips} creates no normalized copy.

\section{Open LLMs}
\label{supp:llm}

Table~\ref{tab:llm} lists the eleven open LLMs of Section~\ref{sec:eval:llm}. Every memory holds the keys of one layer before rotary position embedding (RoPE)~\cite{su2024roformer}, computed on chunks of the PG-19 training books~\cite{rae2020pg19} without their first four positions, which are attention sinks~\cite{xiao2024streamingllm}. The queries of a memory are 10,000 pre-RoPE queries from the held-out test books, drawn uniformly from the query heads of the searched KV head. The middle layer serves where the layers below it fit the four H100 GPUs, and earlier layers serve the largest checkpoints. For multi-head latent attention~\cite{deepseek2024v2}, the memory holds the absorbed key $[\mathbf{c};\mathbf{k}_{\text{rope}}]$ that all heads share, and the query of head $j$ is $[\mathbf{W}_{\text{uk},j}^\top\mathbf{q}_{\text{nope},j};\mathbf{q}_{\text{rope},j}]$. The inner product of key and query then equals the pre-RoPE attention logit. DeepSeek-V4~\cite{deepseek2026v4} shares one key head across all query heads. Each of the four searched heads then takes its queries from one quarter of the query heads. The memory of Qwen3.8-Flash-Next~\cite{qwen2026qwen38} holds 2.5M keys, since its linear-attention layers ran at about 340 tokens per second without fused kernels, and its queries come from 256 positions of each of 40 test chunks. Every compared method reads the same memory of 10M keys per head, or the whole memory where it holds fewer keys.

\begin{table}[t]
\centering
\resizebox{\columnwidth}{!}{
\begin{tabular}{lrrrrr}
\toprule
Model & Layer & Dim & Heads & Chunk & Keys per head \\
\midrule
gpt-oss-120b & 17 of 36 & 64 & 4 & 8,192 & 10.0M \\
Qwen2.5-7B & 14 of 28 & 128 & 4 & 8,192 & 250.0M \\
Qwen2.5-32B & 32 of 64 & 128 & 4 & 8,192 & 40.0M \\
Qwen2.5-72B & 40 of 80 & 128 & 4 & 8,192 & 40.0M \\
Hy3 & 20 of 80 & 128 & 4 & 8,192 & 10.0M \\
Qwen3-235B-A22B & 47 of 94 & 128 & 4 & 8,192 & 10.0M \\
Qwen3.8-Flash-Next & 23 of 48 & 256 & 2 & 8,192 & 2.5M \\
DeepSeek-V4-Flash & 21 of 43 & 512 & 4 & 4,096 & 10.0M \\
DeepSeek-V3.2 & 20 of 61 & 576 & 4 & 4,096 & 10.0M \\
GLM-5.3 & 24 of 78 & 576 & 4 & 4,096 & 10.0M \\
DeepSeek-V4-Pro & 12 of 61 & 512 & 4 & 2,048 & 10.0M \\
\bottomrule
\end{tabular}
}
\caption{Extracted layer and memory of each open LLM.}
\label{tab:llm}
\end{table}

\begin{figure}[t]
\centering
\includegraphics[width=\columnwidth]{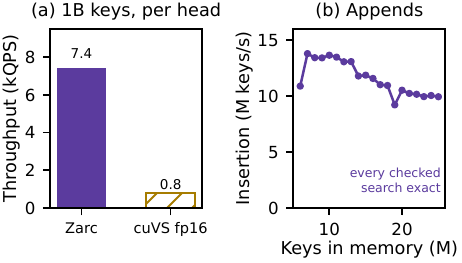}
\caption{Qwen2.5-7B with one billion keys and with appends.}
\label{fig:eval_llm_scale}
\end{figure}

Figure~\ref{fig:eval_llm_scale} extends Qwen2.5-7B~\cite{yang2024qwen25} to one billion keys and to a growing memory. With 250M keys per head, four H100 GPUs hold one billion keys, on which \sysname\ answers 7.4k exact queries per second per head with the raw keys in host memory, 9.5$\times$ exhaustive cuVS~\cite{nvidia2025cuvs} search in half precision (Figure~\ref{fig:eval_llm_scale}(a)). A memory that grows from 5M to 25M keys in steps of 1M absorbs 10.4M to 12.8M keys per second per head while every checked search stays exact (Figure~\ref{fig:eval_llm_scale}(b)).

\section{GPU Generations}
\label{supp:gpus}

\begin{figure}[t]
\centering
\includegraphics[width=\columnwidth]{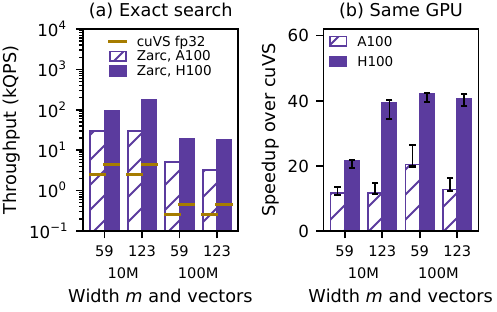}
\caption{Exact throughput on A100 and H100 GPUs.}
\label{fig:eval_gpus}
\end{figure}

We evaluate whether \sysname\ keeps its exactness and its advantage over exhaustive search on an earlier GPU generation. The evaluation compares an NVIDIA A100 with 80~GB of HBM2e memory against the H100 of the other experiments (Section~\ref{sec:eval:setup}), on BIGANN-10M~\cite{jegou2011bigann} and BIGANN-100M at the widths 59 and 123 with batches of 512 to 8,192 queries. The A100 experiments run on a third server with two 64-core AMD EPYC 7763 processors, 503~GB of host DRAM, and one NVIDIA A100 80~GB PCIe GPU, which runs as one MIG instance of the 7g.80gb profile with 98 of its 108 streaming multiprocessors. This server runs the same driver and software stack, except that the library is compiled by CUDA 12.9 with CUTLASS 4.1~\cite{nvidia2025cutlass}. The same rows and thresholds apply on both GPUs, because the rounding certificate bounds the fp32 accumulation error of the product without assuming a summation order. On the H100, \sysname\ answers 86.6k to 98.5k queries per second at $m=59$ and 155k to 181k at $m=123$ on BIGANN-10M over this batch range, against 4.5k for exhaustive cuVS~\cite{nvidia2025cuvs} search. Figure~\ref{fig:eval_gpus} compares the A100 with these values.

On the A100, every query is exact for both sizes, both widths, and all five batch sizes, where every configuration keeps the same survivors per query as on the H100. Figure~\ref{fig:eval_gpus}(a) shows that the A100 answers 28.4k to 37.2k queries per second on BIGANN-10M~\cite{jegou2011bigann} and 3.1k to 6.6k on BIGANN-100M, against 2.5k and 251 for exhaustive cuVS~\cite{nvidia2025cuvs} search on the same GPU. \sysname\ is therefore 11$\times$ to 15$\times$ faster than exhaustive search on BIGANN-10M and 12$\times$ to 26$\times$ faster on BIGANN-100M (Figure~\ref{fig:eval_gpus}(b)), against 19$\times$ to 42$\times$ on the H100. The speedup is smaller on the A100 because the screening pass takes 90\% of the batch time on BIGANN-10M there, and its Ampere kernels, which issue mma.sync instructions, reach a smaller share of the peak tensor-core throughput than the TMA and WGMMA kernels of the H100. For the same reason, the narrower width 59 is the faster one on the A100 at 100M vectors, with 5.0k to 6.6k queries per second, whereas both widths reach the same throughput on the H100.

\section{Evaluation Details}
\label{supp:eval}

\paragraph{Datasets.}
Table~\ref{tab:datasets} summarizes the datasets of Section~\ref{sec:eval:setup} and the workloads that each serves. The evaluation covers local image descriptors, visual foundation-model embeddings, and web-scale multimodal embeddings. SIFT1M and BIGANN~\cite{jegou2011bigann} contain 128-dimensional SIFT~\cite{lowe2004sift} descriptors under Euclidean distance. The BIGANN prefixes from 10M to 1B vectors come with official exact ground truth. DINOv2-1.28M contains 768-dimensional embeddings of the ImageNet-1k~\cite{deng2009imagenet} training images from the DINOv2~\cite{oquab2024dinov2} ViT-B encoder. DataComp-12.8M contains the 12.79 million 768-dimensional CLIP~\cite{radford2021clip} ViT-L/14 image embeddings of the DataComp~\cite{gadre2023datacomp} small pool, normalized to unit length, on which Euclidean distance ranks neighbors as cosine similarity does. Filtered search assigns random labels to BIGANN-10M at 10\%, 1\%, and 0.1\% selectivity. MS MARCO~\cite{nguyen2016msmarco} contains 8.84 million text passages, which we embed with two models. The Sentence-BERT~\cite{reimers2019sbert} model msmarco-distilbert-cos-v5 yields unit vectors for cosine retrieval, while TAS-B~\cite{hofstatter2021tasb} yields unnormalized vectors for inner-product retrieval. Deep1B~\cite{babenko2016deep} contains one billion 96-dimensional float32 image descriptors with 10,000 queries and official ground truth, whose 250M and 500M prefixes take their exact ground truth from a brute force in double precision.

\begin{table}[t]
\centering
\resizebox{\columnwidth}{!}{
\begin{tabular}{lrrll}
\toprule
Dataset & Vectors & Dim & Metric & Workload \\
\midrule
SIFT1M & 1,000,000 & 128 & $L_2$ & Local bounds \\
DINOv2-1.28M & 1,281,167 & 768 & $L_2$ & Local bounds \\
BIGANN & 10M to 1B & 128 & $L_2$ & Tiers, filters, updates, tuning \\
DataComp-12.8M & 12,790,000 & 768 & cosine & 10-NN graph \\
Deep1B & 1,000,000,000 & 96 & $L_2$ & Multi-GPU scaling \\
MS MARCO & 8,841,823 & 768 & cos, IP & Retrieval, inner product \\
\bottomrule
\end{tabular}
}
\caption{Evaluated datasets and workloads.}
\label{tab:datasets}
\end{table}

\paragraph{Ground truth of Deep1B.}
The exact answers also expose three errors in the official ground truth of Deep1B~\cite{babenko2016deep}. On three of the 10,000 queries, the answers of \sysname\ differ from the official ground truth. An independent double-precision brute force over all one billion vectors confirms the answers of \sysname\ in all three cases. Two of these queries have 16,863 and 23,515 exact duplicates at distance zero in the dataset, of which the official list keeps only six. The official 10th neighbor of the third query lies farther than the vector that \sysname\ returns. The Deep1B results consequently use the official ground truth with these three rows corrected, which shows that benchmark ground truth itself needs exact search at a billion vectors.

\section{Exactness Under Adversarial Inputs}
\label{supp:stress}

We test the exactness of \sysname\ on inputs built to break a floating-point bound, which Section~\ref{sec:eval:setup} summarizes. Table~\ref{tab:stress} lists nine inputs of one million 128-dimensional vectors, which range from exact ties and duplicates to values below and far above the normal range of half precision, together with the metrics that each input stresses: Euclidean distance (L2), inner product (IP), and cosine similarity (cos). For every input and metric, 1,000 queries with $k=10$ and $k=100$ are compared with exhaustive search in double precision, where a query counts as exact when its $k$ returned ids are distinct and their double-precision distances equal the $k$ smallest ones.

\begin{table}[t]
\centering
\small
\begin{tabular}{@{}llc@{}}
\toprule
Input & Metrics & Exact queries \\
\midrule
Gaussian vectors & L2, IP, cos & 100\% \\
Rank-4 data & L2 & 100\% \\
50 copies of every vector & L2, IP & 100\% \\
Coordinates in $\{0, 1, 2\}$ & L2, IP & 100\% \\
Pairs one float32 ulp apart & L2 & 100\% \\
Gaussian scaled by $10^{-6}$ & L2, cos & 100\% \\
Gaussian scaled by $10^{4}$ & L2, IP & 100\% \\
Norms $10^{-3}$ and $10^{3}$ mixed & L2, IP & 100\% \\
Gaussian, 10\% deleted & L2 & 100\% \\
\bottomrule
\end{tabular}
\caption{Exact queries under adversarial inputs.}
\label{tab:stress}
\end{table}

Every query is exact on every input, for both values of $k$ and for all 32 combinations of input, metric, and $k$. The inputs exercise the two places where a floating-point search can fail. Ties and duplicates at the $k$-th distance test whether verification in double precision resolves the order that half precision cannot, whereas the scaled and mixed norms test whether the rounding-error bound of every screening row still covers the error of its product.

\section{Semantic Deduplication}
\label{supp:allknn}

\begin{figure}[t]
\centering
\includegraphics[width=\columnwidth]{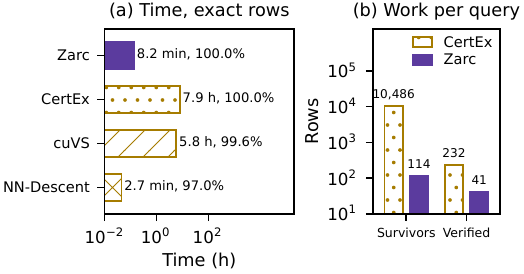}
\caption{The exact 10-NN graph of DataComp-12.8M.}
\label{fig:eval_allknn}
\end{figure}

Figure~\ref{fig:eval_allknn} details the deduplication workload of Section~\ref{sec:eval:apps}, with the end-to-end time and the exact rows of every method for the 10-nearest-neighbor graph of DataComp-12.8M~\cite{gadre2023datacomp}, together with the survivors and the verified rows per query of \sysname\ and of the CertEx engine~\cite{zhao2026certex}. \sysname\ finishes in 8.2 minutes, 42$\times$ faster than the projected 5.8 hours of exhaustive cuVS~\cite{nvidia2025cuvs} search, while NN-Descent~\cite{dong2011nndescent} leaves 3.0\% of the rows inexact (Figure~\ref{fig:eval_allknn}(a)). The partition threshold cuts the survivors per query from 10,486 to 114, of which verification reads 41 (Figure~\ref{fig:eval_allknn}(b)).

\end{document}